\documentclass[journal,comsoc]{IEEEtran}
\usepackage[utf8]{inputenc}
\IEEEoverridecommandlockouts

\usepackage{color,soul}
\usepackage{enumitem}
\usepackage{xcolor}
\usepackage{orcidlink}
\usepackage{colortbl}

\usepackage{verbatim}
\usepackage{multirow}
\usepackage{cite}
\usepackage{tikz}
\usetikzlibrary{arrows.meta}
\usetikzlibrary{positioning}
\usepackage{amsmath,amssymb,amsfonts}
\usepackage{algorithmic}
\usepackage{graphicx}
\usepackage{textcomp}
\usepackage{tabularx}
\usepackage{longtable}
\sethlcolor{yellow} 
\def\BibTeX{{\rm B\kern-.05em{\sc i\kern-.025em b}\kern-.08em
 T\kern-.1667em\lower.7ex\hbox{E}\kern-.125emX}}
\usepackage{footnote}
\makesavenoteenv{tabular}
\makesavenoteenv{table} 
\usepackage{float}
\usepackage{acronym}

\acrodef{3GPP}[3GPP]{Third Generation Partnership Project}
\acrodef{A1AP}[A1AP]{A1 Application Protocol}
\acrodef{A1EI}[A1-EI]{A1 Enrichment Information}
\acrodef{A1TD}[A1TD]{A1 Type Definitions}
\acrodef{A2A}[A2A]{Agent-to-Agent}
\acrodef{AAL}[AAL]{Acceleration Abstraction Layer}
\acrodef{ACP}[ACP]{Agent Communication Protocol}
\acrodef{AI}[AI]{Artificial Intelligence}
\acrodef{AIML}[AI/ML]{Artificial Intelligence/Machine Learning}
\acrodef{AIRAN}[AI-RAN]{Artificial Intelligence Radio Access Network}
\acrodef{AMF}[AMF]{Access and Mobility Management Function}
\acrodef{ANP}[ANP]{Agent Network Protocol}
\acrodef{ANN}[ANN]{Artificial Neural Network}
\acrodef{API}[API]{Application Programming Interface}

\acrodef{BILP}[BILP]{Binary Integer Linear Programming}
\acrodef{BiDLSTM}[BiDLSTM]{Bidirectional Deep Long Short-Term Memory}
\acrodef{BLER}[BLER]{Block Error Rate}
\acrodef{BSR}[BSR]{Buffer Status Report}

\acrodef{CAIF}[CAIF]{Contract-based Agentic Intent Framework}
\acrodef{CAMARA}[CAMARA]{Common API Mobile and Routing Access}
\acrodef{CARC}[CARC]{Channel-Aware Radio Unit ON/OFF Controller}
\acrodef{CDPM}[CDPM]{Central Unit--Distributed Unit Placement Module}
\acrodef{CFR}[CFR]{Channel Frequency Response}
\acrodef{CIR}[CIR]{Channel Impulse Response}
\acrodef{CM}[CM]{Configuration Management}
\acrodef{CNN}[CNN]{Convolutional Neural Network}
\acrodef{CPU}[CPU]{Central Processing Unit}
\acrodef{CQI}[CQI]{Channel Quality Indicator}
\acrodef{CRD}[CRD]{Custom Resource Definition}
\acrodef{CSI}[CSI]{Channel State Information}
\acrodef{CSIRS}[CSI-RS]{Channel State Information Reference Signal}

\acrodef{DAE}[DAE]{Deep Autoencoder}
\acrodef{DDoS}[DDoS]{Distributed Denial of Service}
\acrodef{DID}[DID]{Decentralized Identifier}
\acrodef{DL}[DL]{Deep Learning}
\acrodef{MLDL}[ML/DL]{Machine Learning/Deep Learning}

\acrodef{DMRS}[DM-RS]{Demodulation Reference Signal}
\acrodef{DNN}[DNN]{Deep Neural Network}
\acrodef{DQNMARL}[DQN-MARL]{Deep Q-Network-Multi-Agent Reinforcement Learning}
\acrodef{DRB}[DRB]{Data Radio Bearer}
\acrodef{DT}[DT]{Decision Tree}

\acrodef{E2AP}[E2AP]{E2 Application Protocol}
\acrodef{E2SM}[E2SM]{E2 Service Model}
\acrodef{E2SMCCC}[E2SM-CCC]{E2SM for Cell Configuration and Control}
\acrodef{E2SMDAPP}[E2SM-DAPP]{E2SM for distributed applications}
\acrodef{E2SMKPM}[E2SM-KPM]{E2SM for KPM}
\acrodef{E2SMNI}[E2SM-NI]{E2SM for Network Interface}
\acrodef{E2SMRC}[E2SM-RC]{E2SM for RAN Control}
\acrodef{EI}[EI]{Enrichment Information}
\acrodef{eMBB}[eMBB]{enhanced Mobile Broadband}
\acrodef{eNB}[eNB]{evolved NodeB}
\acrodef{eCPRI}[eCPRI]{enhanced Common Public Radio Interface}
\acrodef{ES}[ES]{Energy Saving}

\acrodef{F1AP}[F1AP]{F1 Application Protocol}
\acrodef{FL}[FL]{Federated Learning}
\acrodef{FM}[FM]{Fault Management}
\acrodef{FTPeS}[FTPeS]{File Transfer Protocol over explicit TLS}

\acrodef{GBR}[GBR]{Guaranteed Bit Rate}
\acrodef{GDPR}[GDPR]{General Data Protection Regulation}
\acrodef{GNN}[GNN]{Graph Neural Network}
\acrodef{gNB}[gNB]{next-generation NodeB}
\acrodef{GPT}[GPT]{Generative Pre-trained Transformer}
\acrodef{GTPU}[GTP-U]{General Packet Radio Service Tunnelling Protocol User Plane}
\acrodef{GPU}[GPU]{Graphics Processing Unit}
\acrodef{HARQ}[HARQ]{Hybrid Automatic Repeat Request}
\acrodef{HTTP}[HTTP]{Hypertext Transfer Protocol}
\acrodef{HTTPS}[HTTPS]{Hypertext Transfer Protocol Secure}

\acrodef{IE}[IE]{Information Element}
\acrodef{IID}[IID]{Independent and Identically Distributed}
\acrodef{NONIID}[non-IID]{non-Independent and Identically Distributed}
\acrodef{IP}[IP]{Internet Protocol}
\acrodef{IQ}[I/Q]{In-phase and Quadrature}
\acrodef{IoT}[IoT]{Internet of Things}

\acrodef{JSON}[JSON]{JavaScript Object Notation}

\acrodef{KNN}[KNN]{K-Nearest Neighbours}
\acrodef{KPI}[KPI]{Key Performance Indicator}
\acrodef{KPM}[KPM]{Key Performance Measurement}

\acrodef{LLM}[LLM]{Large Language Model}
\acrodef{LR}[LR]{Logistic Regression}
\acrodef{LSTM}[LSTM]{Long Short-Term Memory}

\acrodef{MAC}[MAC]{Medium Access Control}
\acrodef{MARL}[MARL]{Multi-Agent Reinforcement Learning}
\acrodef{MCP}[MCP]{Model Context Protocol}
\acrodef{MCS}[MCS]{Modulation and Coding Scheme}
\acrodef{MEC}[MEC]{Multi-access Edge Computing}
\acrodef{MITM}[MITM]{Man-In-The-Middle}
\acrodef{ML}[ML]{Machine Learning}
\acrodef{MLP}[MLP]{MultiLayer Perceptron}
\acrodef{MLOps}[MLOps]{Machine Learning Operations}
\acrodef{mMTC}[mMTC]{massive Machine-Type Communications}

\acrodef{NAS}[NAS]{Non-Access Stratum}
\acrodef{NETCONF}[NETCONF]{Network Configuration Protocol}
\acrodef{NGRAN}[NG-RAN]{Next-Generation Radio Access Network}
\acrodef{NI}[NI]{Network Interface}
\acrodef{NN}[NN]{Neural Network}
\acrodef{NR}[NR]{New Radio}
\acrodef{NearRT}[Near-RT]{Near-Real-Time}
\acrodef{NearRTRIC}[Near-RT RIC]{Near-Real-Time RAN Intelligent Controller}
\acrodef{NonRT}[Non-RT]{Non-Real-Time}
\acrodef{NonRTRIC}[Non-RT RIC]{Non-Real-Time RAN Intelligent Controller}

\acrodef{OAI}[OAI]{OpenAirInterface}
\acrodef{OAM}[OAM]{Operations, Administration, and Maintenance}
\acrodef{OCU}[O-CU]{Open Centralized Unit}
\acrodef{OCUCP}[O-CU-CP]{Open Centralized Unit Control Plane}
\acrodef{OCUUP}[O-CU-UP]{Open Centralized Unit User Plane}
\acrodef{ODU}[O-DU]{Open Distributed Unit}
\acrodef{OFH}[O-FH]{Open Fronthaul}
\acrodef{OFDM}[OFDM]{Orthogonal Frequency-Division Multiplexing}
\acrodef{OPRO}[OPRO]{Optimization by Prompting}
\acrodef{ORAN}[O-RAN]{Open Radio Access Network}
\acrodef{ORU}[O-RU]{Open Radio Unit}

\acrodef{PDU}[PDU]{Protocol Data Unit}
\acrodef{PDCP}[PDCP]{Packet Data Convergence Protocol}
\acrodef{PER}[PER]{Packet Error Rate}
\acrodef{PHY}[PHY]{Physical Layer}
\acrodef{PM}[PM]{Performance Management}
\acrodef{PQC}[PQC]{Post-Quantum Cryptography}
\acrodef{PRACH}[PRACH]{Physical Random Access Channel}
\acrodef{PRB}[PRB]{Physical Resource Block}

\acrodef{QFI}[QFI]{QoS Flow Identifier}
\acrodef{QoE}[QoE]{Quality of Experience}
\acrodef{QoS}[QoS]{Quality of Service}
\acrodef{QoSBRO}[QoS-BRO]{QoS-Based Radio Resource Optimization}
\acrodef{QLoRA}[QLoRA]{Quantized Low-Rank Adaptation}

\acrodef{RAG}[RAG]{Retrieval-Augmented Generation}
\acrodef{RAN}[RAN]{Radio Access Network}
\acrodef{RF}[RF]{Radio Frequency}
\acrodef{RIC}[RIC]{RAN Intelligent Controller}
\acrodef{RLC}[RLC]{Radio Link Control}
\acrodef{RLF}[RLF]{Radio Link Failure}
\acrodef{RL}[RL]{Reinforcement Learning}
\acrodef{RPC}[RPC]{Remote Procedure Call}
\acrodef{RRC}[RRC]{Radio Resource Control}
\acrodef{RSRP}[RSRP]{Reference Signal Received Power}
\acrodef{RSRQ}[RSRQ]{Reference Signal Received Quality}
\acrodef{RSSI}[RSSI]{Received Signal Strength Indicator}
\acrodef{RT}[RT]{Real-Time}

\acrodef{SCTP}[SCTP]{Stream Control Transmission Protocol}
\acrodef{SDAP}[SDAP]{Service Data Adaptation Protocol}
\acrodef{SDR}[SDR]{Software-Defined Radio}
\acrodef{SFTP}[SFTP]{SSH File Transfer Protocol}
\acrodef{SINR}[SINR]{Signal-to-Interference-plus-Noise Ratio}
\acrodef{SLA}[SLA]{Service-Level Agreement}
\acrodef{SLM}[SLM]{Small Language Model}
\acrodef{SMO}[SMO]{Service Management and Orchestration}
\acrodef{SOS}[SOS]{Security Operations Scheduling}
\acrodef{SRS}[SRS]{Sounding Reference Signal}
\acrodef{SSB}[SSB]{Synchronization Signal Block}
\acrodef{SSH}[SSH]{Secure Shell}
\acrodef{STFT}[STFT]{Short-Time Fourier Transform}
\acrodef{SUPI}[SUPI]{Subscription Permanent Identifier}
\acrodef{SVM}[SVM]{Support Vector Machine}

\acrodef{TA}[TA]{Timing Advance}
\acrodef{TCP}[TCP]{Transmission Control Protocol}
\acrodef{TEID}[TEID]{Tunnel Endpoint Identifier}
\acrodef{TLS}[TLS]{Transport Layer Security}
\acrodef{TS}[TS]{Technical Specification}

\acrodef{UDP}[UDP]{User Datagram Protocol}
\acrodef{UE}[UE]{User Equipment}
\acrodef{UL}[UL]{UpLink}
\acrodef{UPF}[UPF]{User Plane Function}
\acrodef{URLLC}[URLLC]{Ultra-Reliable Low-Latency Communications}

\acrodef{ViT}[ViT]{Vision Transformer}

\acrodef{W3C}[W3C]{World Wide Web Consortium}
\acrodef{WPFM}[WPFM]{Wireless Physical-Layer Foundation Model}

\acrodef{XGBoost}[XGBoost]{Extreme Gradient Boosting}

\acrodef{YAML}[YAML]{YAML Ain't Markup Language}

\acrodef{nGRG}[nGRG]{next Generation Research Group}

\acrodef{PCI}[PCI] {Physical Cell Identity}

\acrodef{REST}[REST] {Representational State Transfer}

\acrodef{AdaBoost}[AdaBoost]{Adaptive Boosting}
\acrodef{QDA}[QDA]{Quadratic Discriminant Analysis}

\acrodef{FNN}[FNN]{Feedforward Neural Network}
\acrodef{INT}[INT]{In-band Network Telemetry}
\acrodef{IOAM}[IOAM]{In situ Operations, Administration, and Maintenance}

\acrodef{NCRT}[NCRT]{Neighbor Cell Relation Table}

\acrodef{A1ML}[A1-ML]{A1 ML Model Management Service}

\begin{document}

 \bstctlcite{IEEEexample:BSTcontrol}
 \title{From Telemetry to Intelligence:\\A Data-Centric Survey of O-RAN}

\author{
Samara Mayhoub\,\orcidlink{0000-0001-7629-0532}, Zhengjia Xu\,\orcidlink{0000-0001-5554-6076}, Tareq M. Shami\,\orcidlink{0000-0002-0159-4985}, Mohammad Asif Habibi\,\orcidlink{0000-0001-9874-0047}, Hamed Alimohammadi\,\orcidlink{0000-0001-6005-691X}, Ayhan Akbas\,\orcidlink{0000-0002-6425-104X}, Chuan Heng Foh\,\orcidlink{0000-0002-5716-1396}, and Jose Maria Alcaraz Calero\,\orcidlink{0000-0002-2654-7595}

\par\vspace{2mm}
\parbox{0.95\textwidth}{%
    \centering
    \small
    \textbf{Authors' Note:}
    This is a preprint version of the article currently under review.
    We welcome constructive feedback and comments to improve
    the quality and clarity of this work.
    Please contact the corresponding author,
    Samara Mayhoub (s.mayhoub@aston.ac.uk).

    \par\vspace{3mm}

    \textbf{Disclaimer:}
    This work has been submitted to the IEEE for possible
    publication. Copyright may be transferred without notice,
    after which this version may no longer be accessible.
}
\par\vspace{3mm}

\vspace{-6mm}

\thanks{
S. Mayhoub\orcidlink{0000-0001-7629-0532}, Z. Xu\orcidlink{0000-0001-5554-6076} and J. M. Alcaraz Calero\orcidlink{0000-0002-2654-7595} are with Aston CyberHub- Research Center for Cyber Security, Aston University, Birmingham, UK. T. M. Shami\orcidlink{0000-0002-0159-4985} is with the University of York, Heslington, York, UK. M. A. Habibi\orcidlink{0000-0001-9874-0047} is with Intelligent Networks (IN) Research Department, German Research Center for Artificial Intelligence (DFKI), Kaiserslautern, Germany. H. Alimohammadi\orcidlink{0000-0001-6005-691X} and C. H. Foh\orcidlink{0000-0002-5716-1396} are with the 6G Innovation Centre, Institute for Communication Systems, University of Surrey, Guildford, UK. A. Akbas\orcidlink{0000-0002-6425-104X} is with Computer Engineering Department, Ankara Bilim University, Ankara, Turkiye.
 The corresponding author to whom correspondence should be addressed is S. Mayhoub\orcidlink{0000-0001-7629-0532} (s.mayhoub@aston.ac.uk).
\vspace{-1.4mm}
}

}

\maketitle

\begin{abstract}
The Open Radio Access Network (O-RAN) has emerged as a new architecture for enabling programmability, intelligence, automation, and multi-vendor interoperability in the wireless domain of public telecommunications networks. Artificial Intelligence/Machine Learning (AI/ML), and more recently Agentic AI, are considered key enablers of automation and intelligentization in O-RAN. However, the effectiveness of such systems is constrained by the availability, quality, and structure of O-RAN data. In this survey, we present a data-centric perspective of O-RAN, focusing on how data support rule-based, AI/ML-driven, and emerging Agentic AI applications across various timescale control loops. We analyze O-RAN data along multiple dimensions, including source, accessibility, abstraction level, and application. We further identify key data challenges, and discuss their implications for intelligent control. The overarching objective of this work is to present the first comprehensive review of the data-centric perspective on intelligent applications in O-RAN. To this end, it introduces a unified framework for understanding the role of O-RAN data in enabling network intelligence and automation.
\end{abstract}

\begin{IEEEkeywords}
5G, 6G, Agentic AI, AI/ML, Beyond 5G, dApp, intelligent controller, management data, network data, O-RAN, Open RAN, rApp, service data, xApp.
\end{IEEEkeywords}

\acresetall
\section{Introduction} \label{sec:intro}
\IEEEPARstart{T}{he} \ac{ORAN} represents a shift in the design, operation and the deployment of next-generation mobile networks~\cite{10024837, 11124199}. It adopts four key principles: (i) disaggregation of network components; (ii) virtualization of network functions; (iii) openness of network interfaces; and (iv) intelligent control of network services. All those four principles provide the mobile network with flexibility, scalability, interoperability and intelligence~\cite{10024837}.

Recent research on \ac{ORAN} has focused primarily on three main directions. First, \textit{standardizing and implementing} the \ac{ORAN} concept, its architecture, open interfaces, and deployment strategies. The standardization efforts are mainly led by the \ac{ORAN} Alliance~\cite{O-RAN-Alliance} and \ac{3GPP} specifications~\cite{3GPP}. The software-based implementations are led by different industrial and academic organizations, such as the open-source protocol stacks srsRAN~\cite{10.1145/2980159.2980163} and \ac{OAI}~\cite{10.1145/2677046.2677053}. A tutorial paper on \ac{ORAN} deployment and implementation solutions provides more information in~\cite{11028899}. Second, \textit{investigating the security challenges} introduced by this new, open, disaggregated and programmable \ac{RAN}. These challenges include threats against open interfaces, \ac{RAN}, controllers, as well as risks arising from malicious applications, data poisoning, misconfiguration, and unauthorized control actions~\cite{11370849, 10613612, 10467183,10339923, 11250611, 10606000}. Third, \textit{proposing network management applications} and deploying them on the \ac{ORAN} architecture, following the app-store-like model where network functions can be extended through third-party software~\cite{11262686}. Examples of these applications include network slicing, traffic steering, interference mitigation, anomaly detection, and energy optimization, operating across different time scales (\ac{RT}, \ac{NearRT}, and \ac{NonRT}). In these applications, \ac{AIML} models are widely adopted, and recent work is also moving toward Agentic \ac{AI}, where \ac{LLM}-based or multi-agent systems can interpret intent and reason over \ac{ORAN} data, call tools, and coordinate control actions.
However, both intelligent and Agentic \ac{AI}-based \ac{ORAN} applications are fundamentally constrained by the data that are generated, exposed, and accessible across architectural components. While existing studies demonstrate the potential of \ac{AIML} and Agentic \ac{AI} in \ac{ORAN}, they largely concentrate on algorithmic innovation (the methods) rather than examining the underlying data foundations that enable such intelligence. This paper addresses this gap by introducing a data-centric perspective of \ac{ORAN}.

Moreover, the transition to Agentic \ac{AI} makes the data-centric perspective even more important, as agentic \ac{ORAN} do not only require numerical data for model inference, but also need heterogeneous inputs such as intent, logs, policies, standards, tool descriptions, historical decisions, and feedback. Therefore, the reliability of agentic control depends strongly on what data are exposed, how data are abstracted, and whether data are trustworthy and timely. In addition, this trend of \ac{AI} and Agentic \ac{AI} applications is further reinforced by the emerging \ac{AIRAN} vision, in which \ac{AI} is used not only to optimize \ac{RAN} operation but also to support \ac{AI} workloads on the \ac{RAN} components~\cite{10360202,11474793}. This direction is also aligned with early 6G standardization activities~\cite{3GPPRel20}.

\begin{table*}[t]
\centering
\renewcommand{\arraystretch}{1.6}
\caption{Providing a comprehensive comparison between the proposed work and recent state-of-the-art \ac{ORAN} literature across the major technical aspects captured in the corresponding columns}
\label{tab:related_compare_dimensions}
\begin{tabular}{|p{1.3cm}|p{6cm}|c|c|c|c|c|}
\hline
\rowcolor{lightgray} ~~\textbf{Work} & ~~~~~~~~~~~~~~~~~~~ \textbf{Primary Focus} & \textbf{Architecture} & \textbf{Security Challenges} & \textbf{Applications} & \textbf{Agentic \ac{AI}} & \textbf{Data} \\
\hline

~\cite{10024837} &
\acs{ORAN} architecture, interfaces, intelligent control workflows, and security &
\checkmark & \checkmark & \checkmark & -- & -- \\
\hline

~\cite{11124199} &
\acs{ORAN} network slicing architecture, deployment, and orchestration &
\checkmark & -- & \checkmark & -- & -- \\
\hline

~\cite{11028899} &
\acs{ORAN} Application development and deployment across simulation, emulation, and real testbeds &
\checkmark & -- & \checkmark & -- & -- \\
\hline

~\cite{11370849} & ~
\acs{ML} techniques for \acs{ORAN} applications in spectrum management, resource allocation, and security &
\checkmark & \checkmark & \checkmark & -- & -- \\
\hline

~\cite{10877797} &
End-to-end xApp development, lifecycle management, \acs{RAN} control, and testing&
-- & -- & \checkmark & -- & -- \\
\hline

~\cite{10.1145/3777379} &
Intelligence in \acs{ORAN}: \acs{AIML} integration, applications, challenges, and impacts&
-- & -- & \checkmark & -- & -- \\
\hline

~\cite{liyanage_open_2023} &
\ac{ORAN} security and privacy threats, vulnerabilities, and mitigation &
-- & \checkmark & -- & -- & -- \\
\hline

~\cite{10508126} &
Security of open \acs{ORAN} interfaces : E2 and \acs{OFH}&
-- & \checkmark & -- & -- & -- \\
\hline

~\cite{10651559} &
xApp architecture, security vulnerabilities, and mitigation in \acs{ORAN} &
-- & \checkmark & \checkmark & -- & -- \\
\hline

~\cite{s23218792} &~\acs{ML}-enabled network automation in \acs{ORAN}: applications and challenges&
\checkmark & \checkmark & \checkmark & -- & -- \\
\hline

\textbf{This paper} &
\textbf{Data-centric \ac{ORAN}} &
\textbf{\checkmark} & 
\textbf{\checkmark} & 
\textbf{\checkmark} & 
\textbf{\checkmark} & 
\textbf{\checkmark} \\
\hline

\end{tabular}
\end{table*}

Unlike others survey papers, such as~\cite{10024837, 10877797, 11028899, 11370849, 10.1145/3777379, liyanage_open_2023, 10508126, 10651559, s23218792 } which focus primarily on \ac{ORAN} tutorial, its architecture and implementation, applications or security vulnerabilities, (see Table~\ref{tab:related_compare_dimensions}), we examine \ac{ORAN} through lens of data source, abstraction, accessibility, application and challenges. Moreover, this paper is the first to connect data-centric design with Agentic \ac{AI} control loops in \ac{ORAN}. The key contributions of the paper are summarized as follows:

\begin{itemize}
 
 \item Propose a data-centric framework for \ac{ORAN} that characterizes data according to five complementary views: source, accessibility, abstraction level, application use, and challenges.
 
 \item Map rule-based, \ac{AIML}, and Agentic \ac{AI} based \ac{ORAN} applications to their data dependencies across various control loops.
 
 \item Identify data challenges that affect intelligent and agentic \ac{ORAN} applications.
 
\end{itemize}

\begin{figure*}[t]
 \centering
 \includegraphics[width=0.9 \textwidth]{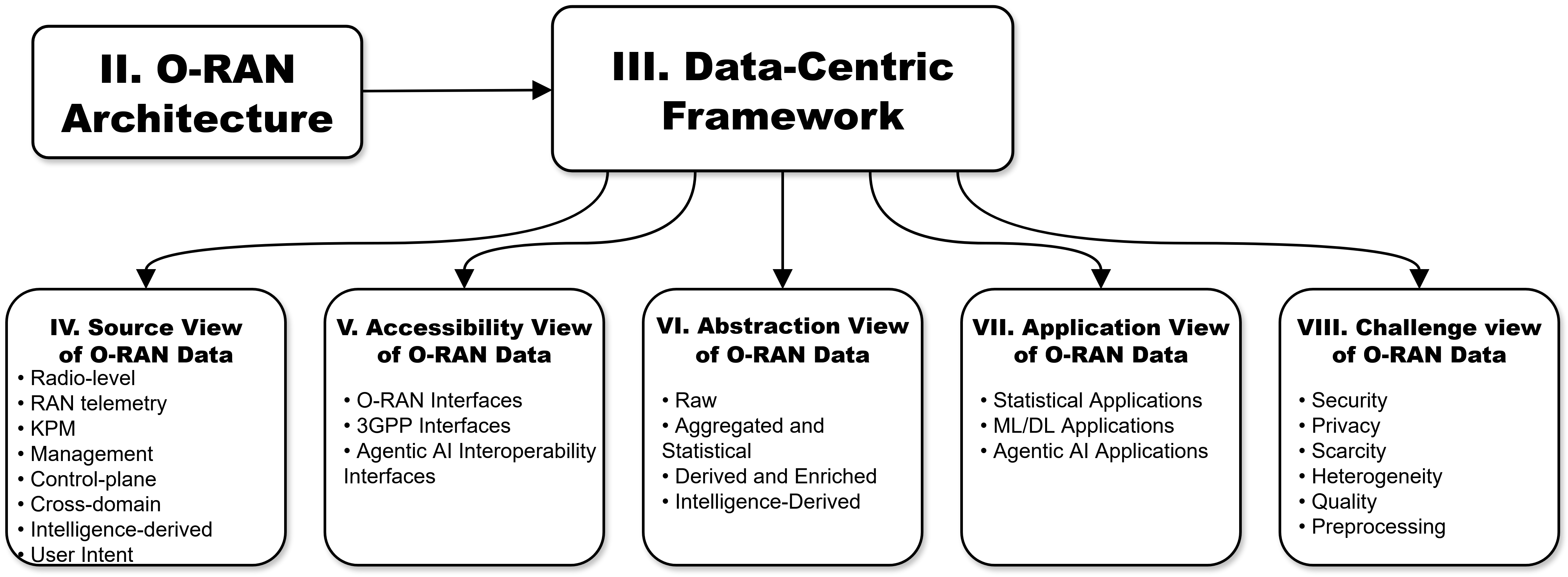}
 \caption{Survey organization:Section~\ref{sec:Architecture} covers \ac{ORAN} architecture. Sections~\ref{sec:framework}--\ref{sec:Abstraction} present the data-centric framework and its source, accessibility, and abstraction views. Section~\ref{sec:Applications} reviews traditional, \ac{AIML} based, and Agentic \ac{AI} applications in \ac{ORAN}. Section~\ref{sec:Challenges} discusses key data challenges.}
 \label{fig:paper_organization}
\end{figure*}

The rest of the paper is organized as follows\footnote{Readers seeking a tutorial on \ac{ORAN} data may read Sections~\ref{sec:SourceDomain}--\ref{sec:Abstraction}, while those mainly interested in intelligent \ac{ORAN} applications may proceed directly to Section~\ref{sec:Applications} and refer back as needed.} (see Fig.~\ref{fig:paper_organization}). Section~\ref{sec:Architecture} introduces \ac{ORAN} architecture. In Section~\ref{sec:framework}, the proposed data-centric framework and its five views: source, accessibility, abstraction level, application, and challenge are presented. Section~\ref{sec:SourceDomain} discusses the source view of \ac{ORAN} data and classifies main data sources and types while Section~\ref{sec:Accessibility} examines the accessibility view by describing \ac{ORAN}, \ac{3GPP}, and Agentic \ac{AI} interfaces through which data are exposed and actions are sent. In Section~\ref{sec:Abstraction}, the abstraction-level view of \ac{ORAN} data is presented. Section~\ref{sec:Applications} reviews traditional, \ac{AIML}-based, and Agentic \ac{AI} \ac{ORAN} applications and maps them to their data dependencies and control loops whereas Section~\ref{sec:Challenges} discusses the main data challenges. Finally, Section~\ref{sec:conclusion} concludes the paper and outlines future research directions.

\color{black}

\section{Background on Key Concepts Related to the \ac{ORAN} Architecture} \label{sec:Architecture} 
The next generation of mobile networks, including beyond-5G and 6G systems, aims to provide an open, flexible, and efficient environment. These networks will support heterogeneous, multi-vendor components. They will also integrate intelligent capabilities to improve the performance of network functions through \ac{AIML} first or third-party applications. This has been realized through applying a new architecture, widely known in industry as \ac{ORAN}. This architecture is currently being standardized by \ac{ORAN} Alliance~\cite{O-RAN-Alliance}.

\ac{ORAN} architecture applies four key principles: disaggregation, virtualization, open interfaces, and \ac{AI}-driven intelligence~\cite{10024837}. These main principles are further discussed below: 
\begin{itemize}

\item \textit{Disaggregation of network components to increase flexibility}: By applying this principle, \ac{ORAN} disaggregates the functionality of the base station component, based on 7.2x Split~\cite{9579445}, into physical \ac{ORU} and virtualized \ac{ODU} and \ac{OCU}. The \ac{OCU} is further divided into \ac{OCUCP} and \ac{OCUUP}. This leaves only \acp{ORU} deployed at the cell site~\cite{10024837}. \acp{ORU} implement the radio frequency and the lower \ac{PHY} functions. On the other hand, \acp{ODU} implement the higher \ac{PHY} functions, \ac{MAC} layer functions and \ac{RLC} layer functions. The \acp{OCU} handle \ac{RRC} layer functions and \ac{PDCP} functions.

While this disaggregation improves flexibility, it also introduces new operational challenges in maintaining consistent state across procedures that span multiple components and open interfaces~\cite{10.1145/3737900.3770170}.

\item \textit{Intelligent controllers to improve the performance of the network}: \ac{ORAN} divides the intelligence of network into two logical \acp{RIC}: \ac{NearRTRIC} and \ac{NonRTRIC}. Both \ac{NearRTRIC} and \ac{NonRTRIC} are expected to leverage data-driven \ac{AIML} algorithms to optimize various network management applications~\cite{10750758}. These intelligent applications can be hosted as extended application (xApps) on the \ac{NearRTRIC} for \ac{NearRT} tasks, and as \ac{RAN} applications (rApps) on the \ac{NonRTRIC} for \ac{NonRT} tasks. Beside xApps and rApps, some intelligent application for real time tasks are deployed as distributed applications (dApp) on \ac{OCU}/\ac{ODU} nodes~\cite{LACAVA2025111342, 9846950} (see also the research report~\cite{nGRG_RR_2024_10_dApps} from \ac{ORAN} \ac{nGRG}.

\item \textit{Open interfaces to allow network components from different vendors to interoperate}: \ac{ORAN} utilizes open interfaces between its network components~\cite{O-RAN.WG1.TS.OAD, O-RAN.WG10.TS.OAM}, such as the \ac{OFH} between \ac{ORU} and \ac{ODU}. The \acp{ODU} and \acp{OCU} are called E2 nodes because they are connected with \ac{NearRTRIC} through the open interface E2~\cite{O-RAN.WG3.TS.E2GAP}. \ac{NonRTRIC} is connected to \ac{NearRTRIC} through the open interface A1~\cite{O-RAN.WG2.TS.A1GAP}. O2 interface is connected \ac{SMO} and O-Cloud~\cite{O-RAN.WG6.TS.O2-GAP}. The open interface O1 is used to connect \ac{SMO} with other \ac{ORAN} logical components including \ac{NearRTRIC} and \ac{ODU}/\ac{OCU}. Moreover, R1 interface is placed between rApps and the \ac{NonRTRIC}~\cite{O-RAN.WG2.TS.R1GAP}.

\item \textit{Virtualization of network functions to enable scalability}: Virtualization in \ac{ORAN} is enabled by O-Cloud, a cloud computing platform (edge and/or central) that provides the physical infrastructure plus the virtualization infrastructure needed to host \ac{ORAN} network functions. Based on \ac{ORAN} architecture description \ac{ORAN} Alliance specifications~\cite{O-RAN.WG1.TS.OAD}, \ac{ORAN} network functions such as \ac{NearRTRIC}, \ac{OCUCP}/\ac{OCUUP}, and \ac{ODU} may be virtualized on the O-Cloud. Moreover, O-Cloud can expose \ac{AAL} \acp{API}~\cite{O-RAN.WG6.AAL} to allow these virtualized components to use hardware accelerators through a common \acp{API} for offloading of compute-intensive physical functions (such as channel coding/decoding)~\cite{ O-RAN.WG6.AAL-GAnP}. Those efforts of Virtualization support the scalability of the mobile network.
\end{itemize}

\begin{figure*}[h]
 \centering
 \includegraphics[width=0.9\textwidth]{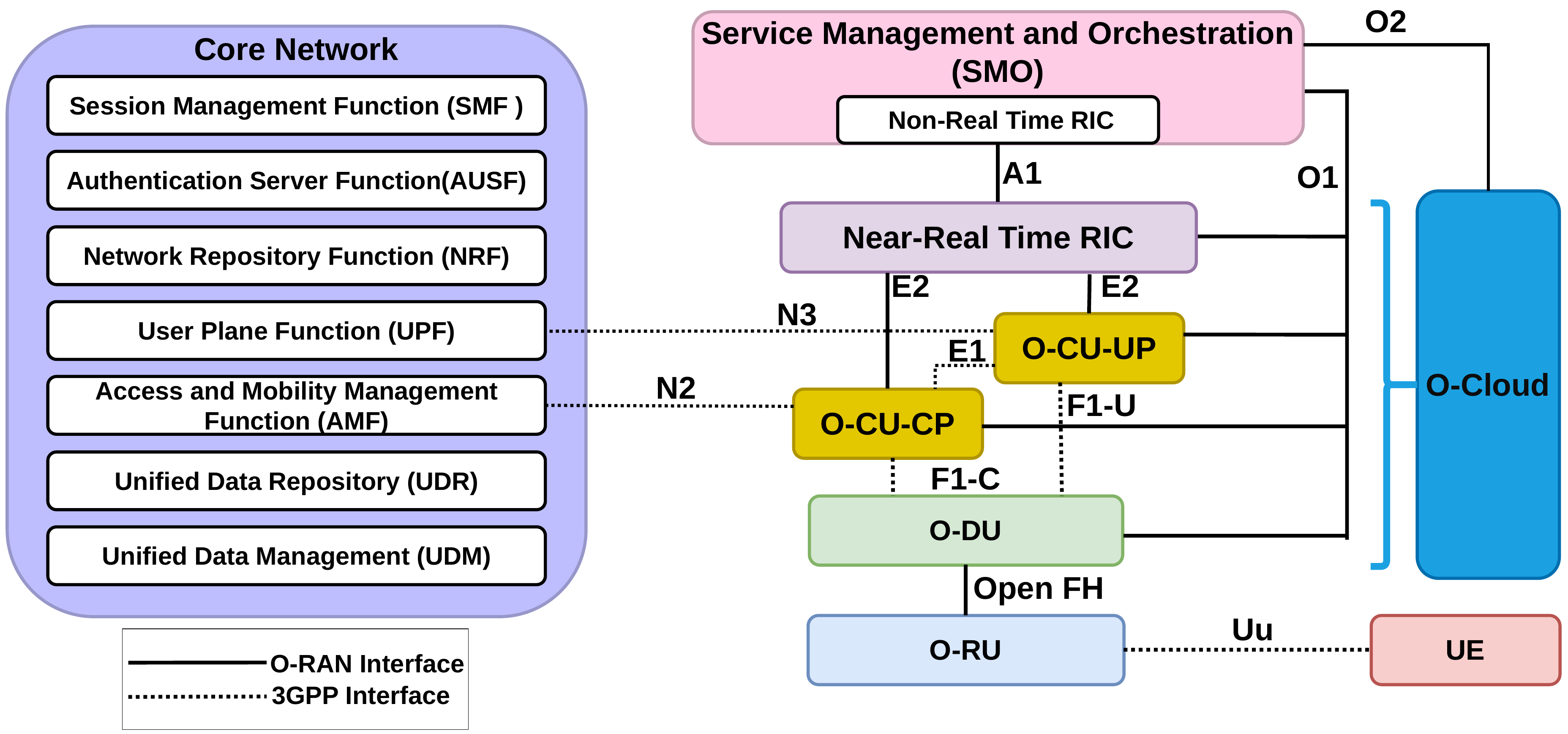}
 \caption{Overview of the \ac{ORAN} architecture and its main interfaces}
 \label{fig:oran_arch_interfaces}
\end{figure*}

Figure~\ref{fig:oran_arch_interfaces} illustrates the main components of \ac{ORAN} and key interfaces among \ac{SMO}, \acp{RIC}, disaggregated \ac{RAN} units, \acp{UE}, and core network components.

\ac{ORAN} supports different classes of software applications deployed at different control points of the architecture and operating at different timescales. The main application classes are rApps, hosted by the \ac{NonRTRIC}, and xApps, hosted by the \ac{NearRTRIC}. In addition, recent research has introduced dApps to support \ac{RT} control directly at \ac{ODU}/\ac{OCU} nodes~\cite{10024837,9846950}. Table~\ref{tab:oran_app_comparison} summarizes the main differences among these application classes.
\begin{itemize}
\item\textit{rApps} are applications executed on top of the \ac{NonRTRIC}, which operates within the \ac{SMO} framework and targets \ac{NonRT} control loops (typically above 1~s). Their role is to see the whole picture and exploit broader network visibility and longer-timescale data, such as collected management data via O1 interface, historical observations, service-level context, and external inputs, in order to perform optimization, policy generation, orchestration, and \ac{AIML} model training management of xApps~\cite{9839628}. Through A1 interface, the\ac{NonRTRIC} and its rApps can provide policy guidance and \ac{EI} to the \ac{NearRTRIC}. Thus, rApps are viewed as higher-level intelligence applications that support long-timescale network optimization and automation.

\item\textit{xApps} are applications executed on the \ac{NearRTRIC}. They are designed for \ac{NearRT} control loops (typically between 10~ms and 1~s). xApps mostly use telemetry (called \ac{KPM}) collected by E2 nodes from \ac{RAN} and delivered through E2 interface by using E2 report procedures and service models such as \ac{E2SMKPM}~\cite{10439167,O-RAN.WG3.E2SM-KPM}. These \acp{KPM} provide valuable insights into the performance of the \ac{RAN} and support data-driven decision making for network optimization~\cite{Systemlevel,Hassouna2025}. In addition, xApps may use policies and \ac{EI} received from rApps through the A1 interface. The outputs of xApps are control decisions and actions sent through E2 to optimize \ac{RAN}.

\item\textit{dApps} are an emerging extension (distributed applications), proposed in the research literature to support \ac{RT} \ac{RAN} control loops that are stricter than those currently addressed by xApps and rApps. Unlike xApps and rApps, which are associated with \acp{RIC}, dApps are deployed directly close to the execution points in \ac{RAN}, typically on or near \ac{ODU} and \ac{OCU} nodes. Their motivation is to process highly local and latency-sensitive data without sending all such data to a remote \ac{RIC}. For this reason, dApps are particularly relevant when privacy, bandwidth, or latency constraints make centralized processing less practical~\cite{9846950}. In this sense, dApps complement xApps and rApps by enabling fine-grained local intelligence at \ac{RT} timescales.
\end{itemize}

\begin{table*}[h!]
\centering

\renewcommand{\arraystretch}{1.6}
\caption{Comparison of \ac{ORAN} application classes in terms of deployment, timescale, used data, and purpose}
\label{tab:oran_app_comparison}
\begin{tabular}{|p{1cm}|p{2cm}|p{4.3cm}|p{4.3cm}|p{4.0cm}|}
\hline
\rowcolor{lightgray}
\centering\textbf{App} ~~~ &
\centering\textbf{Deployment} &
\centering\textbf{Timescale} &
\centering\textbf{Used data} &
\centering\textbf{Purpose}
\tabularnewline
\hline

rApp &
\ac{NonRTRIC} &
\ac{NonRT} control loops ($> 1$ s) &
Network-wide data, historical observations, and management information &
Long-term optimization, orchestration, and policy control
\tabularnewline
\hline

xApp &
\ac{NearRTRIC}&
\ac{NearRT} control loops ($10$ ms -- $1$ s) &
\ac{NearRT} telemetry, and policies &
\ac{NearRT} resource management and network optimization
\tabularnewline
\hline

dApp &
\ac{ODU}/\ac{OCU} &
\ac{RT} control loops ($< 10$ ms) &
Radio data, local host-level runtime data &
\ac{RT} control and optimization on Node-Level
\tabularnewline
\hline

\end{tabular}
\end{table*}

\section{Data-Centric Framework}
\label{sec:framework}
\ac{ORAN} data are inherently heterogeneous as it is generated across disaggregated components, exposed through different interfaces, represented at different abstraction levels, and consumed by different timescale applications. This heterogeneity becomes even more with the emergence of Agentic \ac{AI} applications in \ac{ORAN}, where agents also require semantic data for reasoning.

To organize this data landscape, we introduce a data-centric framework based on five complementary views that capture where \textit{\ac{ORAN} data} originate, how they become accessible, in what form they appear, which applications consume them, and what challenges affect them. Together, they provide a unified structure for understanding the role of data in \ac{ORAN} applications.

\begin{itemize}
\item \textit{Source view:}
This view describes \emph{where the data originate} in \ac{ORAN}. Data may come from different parts of \ac{ORAN} ecosystem, including radio access components, \ac{RAN} functions, management systems, control-plane procedures, core-network and external systems, \ac{ORAN} applications, and Agentic \ac{AI} interactions (such as user intent and feedback).

\item\textit{Accessibility view:}
This view describes \emph{how and through which access path the data can be collected}. In \ac{ORAN}, data accessibility depends on the interface through which data are exposed, this can be a native \ac{ORAN} interface (such as E2, O1, A1, O2), \ac{3GPP} interface (such as N2, N3), or local implementation -specific interface.

\item \textit{Abstraction view:}
This view describes \emph{the level of processing applied to data before they are consumed}. \ac{ORAN} data may appear as raw unprocessed measurements, aggregated indicators summarized over time or entities, statistical summaries, derived features computed from primary data, enriched context obtained from management systems or external sources, or intelligence-derived outputs such as predictions, classifications, or control decisions.

\item \textit{Application view:} This view describes \emph{which class of applications consumes the data}. Based on their underlying method, \ac{ORAN} applications are categorized into traditional applications, \ac{AI}-driven applications, or Agentic \ac{AI} applications. Each of these applications relies on different types of data depending on its objective, timescale, and control role.

\item \textit{Challenge view:} 
This view describes \emph{the main challenges that affect \ac{ORAN} data throughout their lifecycle}, including collection, processing, and consumption. These challenges include data security, privacy, ambiguity, scarcity, heterogeneity, quality, preprocessing and timeliness.
\end{itemize}

The remainder of this paper is structured around the five views mentioned above. We first examine \ac{ORAN} data by source domain, then discuss how such data become accessible through \ac{ORAN}, \ac{3GPP}, or Agentic \ac{AI} interaction interfaces, followed by the abstraction levels at which data appear, before relating these data to traditional, \ac{AIML}-driven, and Agentic \ac{AI} applications, where \ac{ORAN} data are consumed. Finally, the challenge view is presented.

\begin{figure*}[h]
\centering
\includegraphics[width=0.9\textwidth]{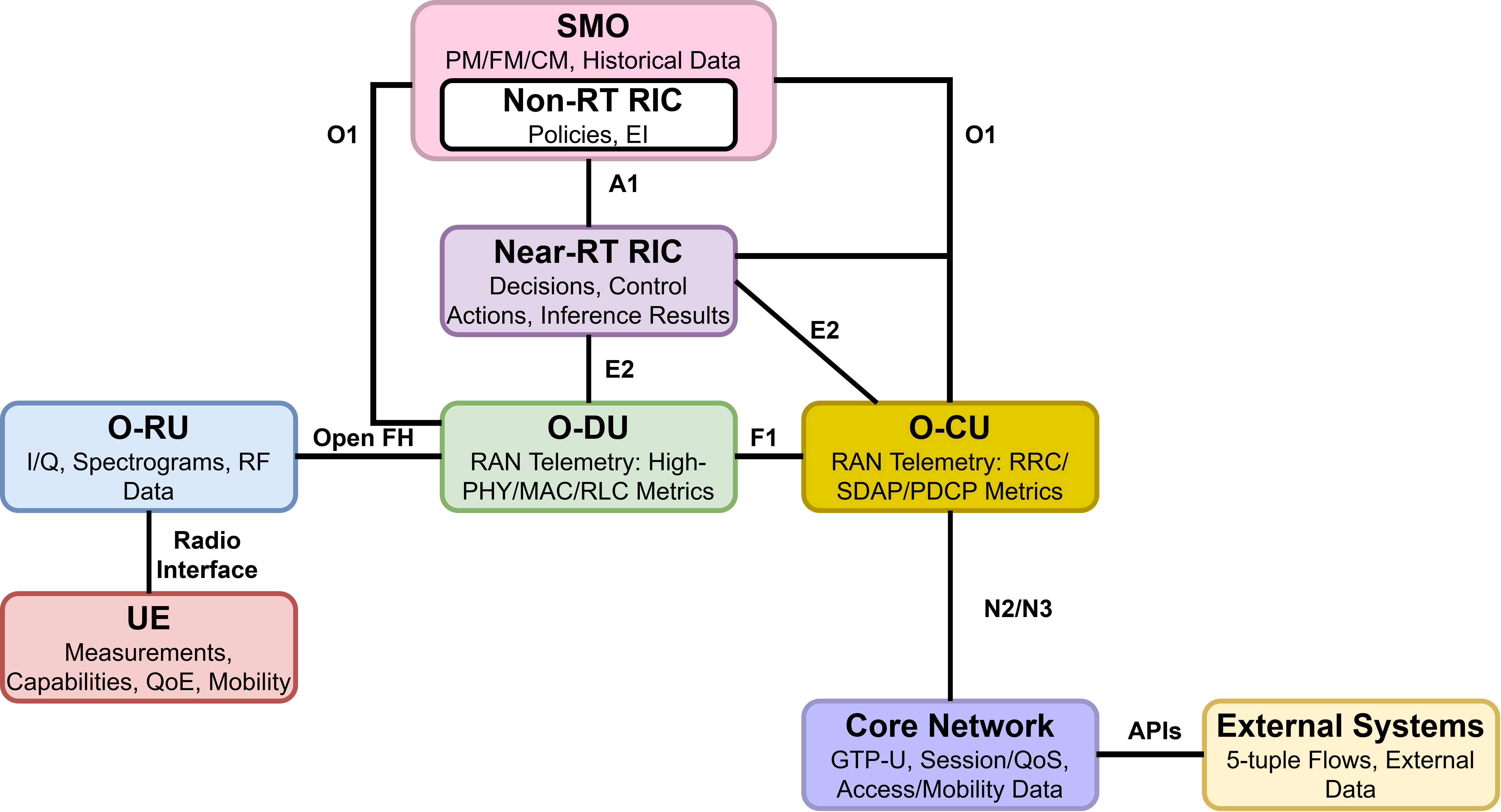}
\caption{ \ac{ORAN} architecture with representative data classes associated with each main entity and the main interfaces used for exposure, control, and integration}
\label{fig:oran_data_architecture}
\end{figure*}

\section{Source View of \ac{ORAN} Data}
\label{sec:SourceDomain}

\ac{ORAN} applications use different types of data, including radio-domain data, \ac{RAN} telemetry, management data, core-network data, and intelligence-derived data. Fig.~\ref{fig:oran_data_architecture} provides an architectural view of the main \ac{ORAN} components and the representative data classes associated with them.
As shown in Fig.~\ref{fig:oran_data_architecture}, different \ac{ORAN} components generate different types of data. \ac{UE} and \ac{ORU} are associated with raw data such as \ac{IQ} samples or \ac{RF} measurements, while \ac{ODU} and \ac{OCU} aggregate and generate more structured data such as \acp{KPM}. Besides \ac{RAN}, the two controllers, \ac{NearRTRIC} and \ac{NonRTRIC}, produce higher-level intelligence data, such as decisions and policies, generated by the applications running on these controllers. On the other hand, historical network data and \ac{PM}, \ac{FM}, and \ac{CM} data are generated at \ac{SMO} component. Since \ac{SMO} has a global view of the whole network, this component is usually used for training \ac{AIML} models and then deploying them at the appropriate component (\ac{NonRTRIC}, \ac{NearRTRIC} or even on \ac{OCU}/\ac{ODU} nodes). The core network generates session/services related data in addition to access/ mobility data of the network components (\acp{UE} and E2 nodes). External systems further contribute traffic, and analytics data generated by external applications such as monitoring systems; these usually contain cross-domain and higher-level data. 

After giving this high-level view of data types, the next subsections explain these data types in more detail with examples from recent developed applications\footnote{Here, application refers to a software application deployed on \ac{ORAN} components, as described in Section~\ref{sec:Architecture}.}.

\subsection{Radio-level data}
Radio-level data capture low-level wireless signal observations. It is mainly produced at the \ac{UE}, \ac{ORU}, and lower \ac{PHY} functions. The following subsections discuss three representative forms: \ac{IQ} samples, spectrograms, and reference signals.

\subsubsection{\ac{IQ} samples} 
\begin{figure}[t]
 \centering
 \includegraphics[width=0.45\textwidth]{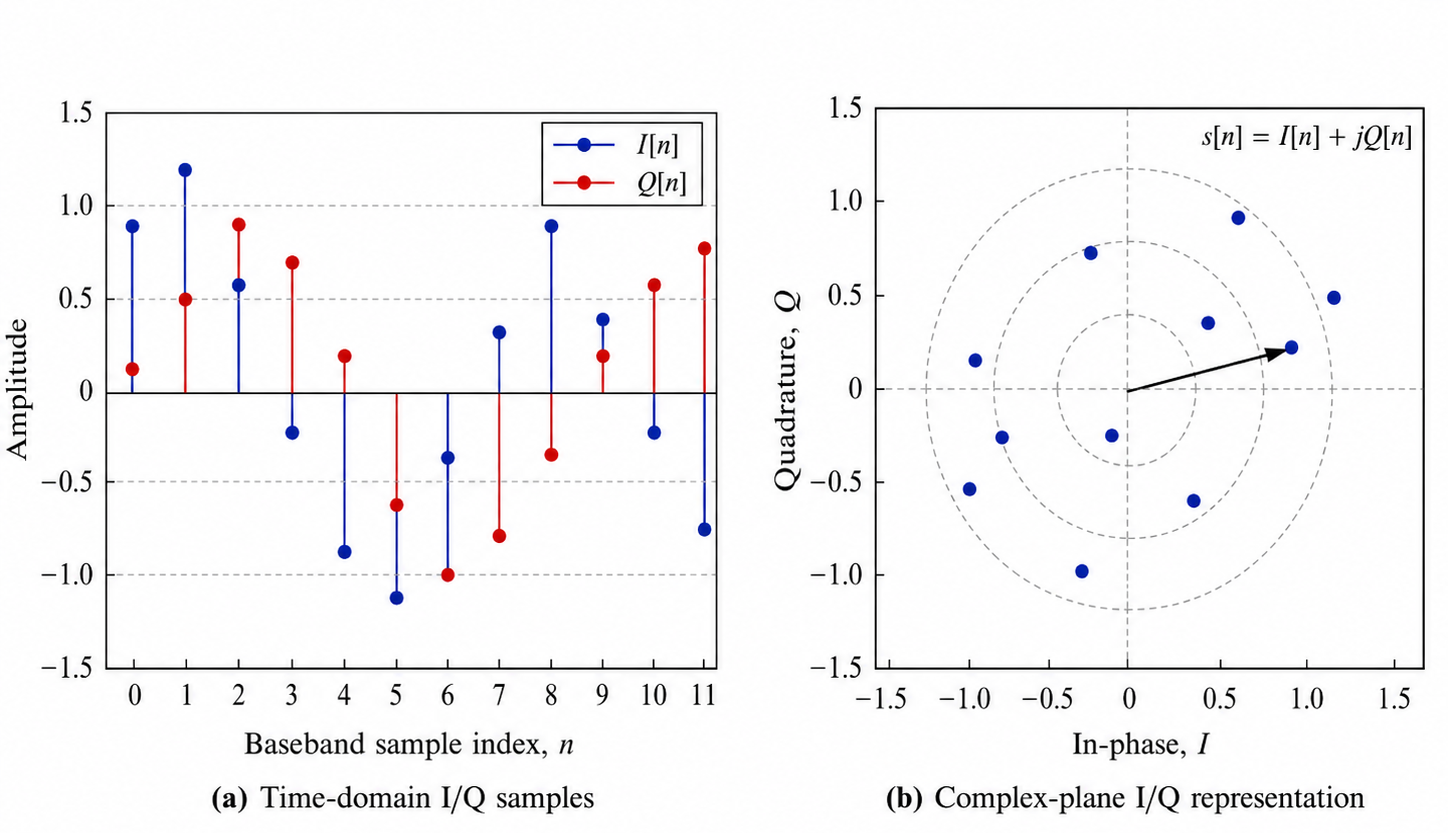}
 \caption{Example representation of \ac{IQ} samples}
 \label{fig:iq_example}
\end{figure}

\ac{IQ} (In-phase and Quadrature) samples are the fundamental representation of wireless \ac{RF} signals, capturing both amplitude and phase information~\cite{barker2025deepsense}. Each \ac{IQ} sample consists of two orthogonal components Fig.~\ref{fig:iq_example}: i) \textit{In-phase (I)} component, which represents the amplitude of the baseband signal that is in-phase with the I carrier (0-degree phase), and ii) \textit{Quadrature (Q)} component, which represents the amplitude of the baseband signal that is in-phase with the Q carrier (90-degree phase).

In the context of \ac{ORAN}, digitized baseband \ac{IQ} samples are exchanged as User-plane (U-plane) traffic over the \ac{OFH} interface between the \ac{ORU} and \ac{ODU}~\cite{O-RAN.WG4.CUS,299752}. This transport commonly relies on Ethernet-based \ac{eCPRI} encapsulation~\cite{eCPRI}. In the \ac{ORAN} U-plane, \ac{IQ} samples are carried in packet payloads together with \ac{ORAN} application headers, and are associated with specific radio resources, such as \ac{OFDM} symbols, resource blocks, antennas, or beams, depending on the system configuration.

The fact that \ac{IQ} data capture the full amplitude and phase information of \ac{RF} signals makes it high-fidelity data~\cite{barker2025deepsense}. This high-fidelity data are crucial for various advanced processing tasks in wireless communications enabling capabilities such as: Spectrum sensing, classification, Interference detection, localization/mitigation, rogue base station detection, \ac{RF} fingerprinting, beam management, dynamic resource allocation and management, anomaly detection, and more. 

It is worth noting that while \ac{IQ} samples are vital for various tasks, transferring high volumes of \ac{IQ} samples directly to \acp{RIC} can saturate network and raise privacy concerns due to containing user information, which often leads to them being processed directly at the base station~\cite{299752}. For this reason, some applications instead send spectrogram images (discussed in the section below) derived from \ac{IQ} samples to the \ac{NearRTRIC}, which can resolve the privacy concern since spectrograms do not allow data decoding~\cite{10356330}.

\subsubsection{Spectrograms} 
\begin{figure}[t]
 \centering
 \includegraphics[width=0.43\textwidth]{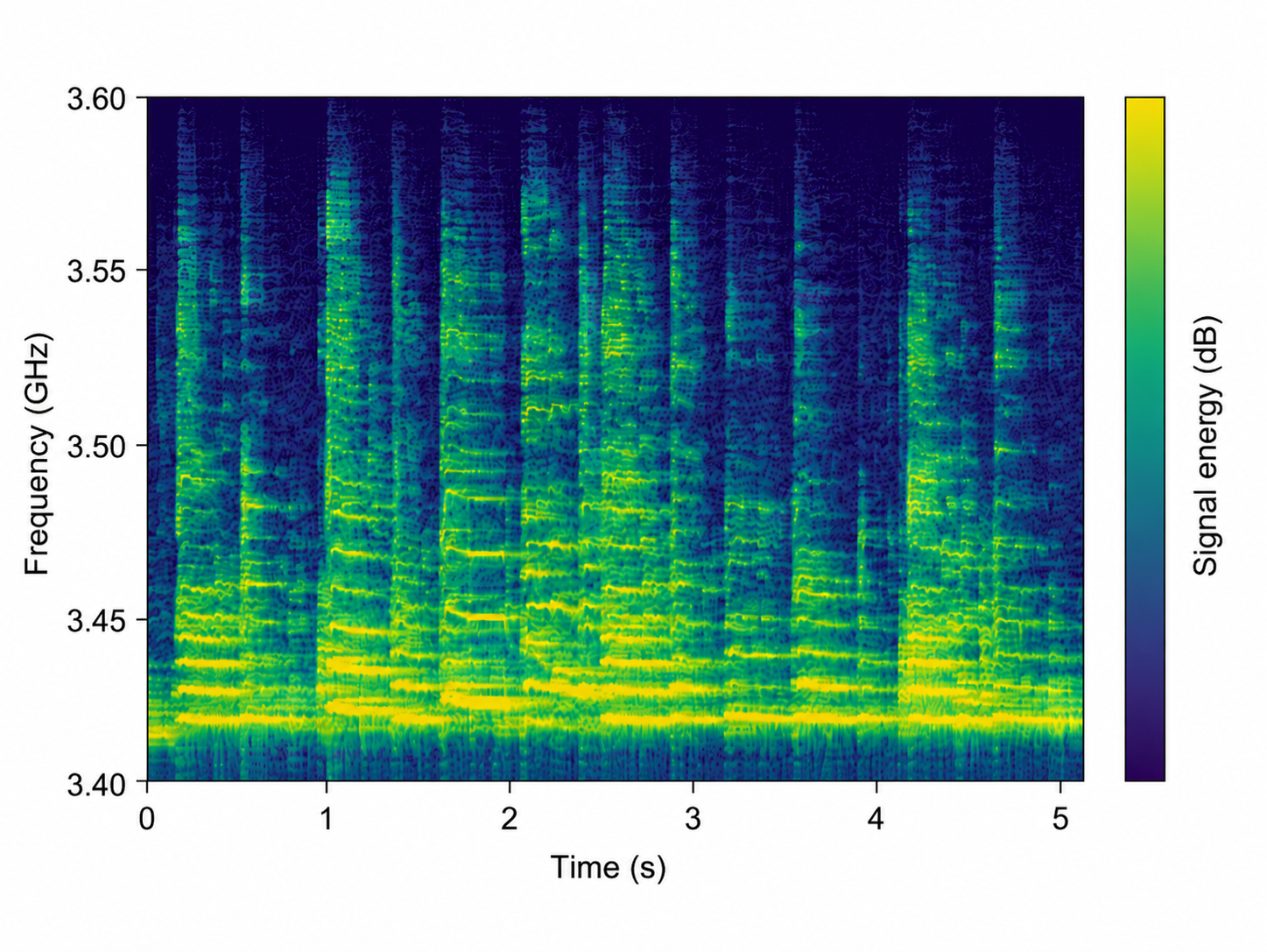}
 \caption{Illustrative spectrogram of wireless signal energy over time and frequency}
 \label{fig:spectrogram_example}
\end{figure}

is a visual (time, frequency) representation that shows how the energy of a signal is distributed across frequency and time~\cite{LAMPERT201087}. In a spectrogram image (see an example at Fig.~\ref{fig:spectrogram_example}) the horizontal axis represents time, the vertical axis represents frequency, and the color usually represents the signal energy in dB at each (time, frequency) point~\cite{10.1155/2024/3376733}. A spectrogram is commonly derived from \ac{IQ} samples by applying \ac{STFT}~\cite{11314504,zou2026rf}.

Compared with raw \ac{IQ} samples (see Table~\ref{tab:iq_vs_spectrogram}), spectrograms provide a more compact representation of wireless signals, which makes spectrograms useful for applications such as identifying interference, and other visually distinguishable signal characteristics. Spectrograms are easier to store and process than raw \ac{IQ} data and can be naturally used by image-based learning models, such as \ac{ViT}~\cite{dosovitskiy2020image}. However, because spectrograms are derived data, they do not preserve the full information of the original signal (such as phase) and are therefore less suitable for tasks requiring detailed signal analysis.

\begin{table}[t]
\centering
\caption{Comparison of raw \ac{IQ} samples and spectrograms}
\label{tab:iq_vs_spectrogram}
\renewcommand{\arraystretch}{1.6}

\begin{tabular}{|p{1.7cm}|p{2.95cm}|p{2.95cm}|}
\hline
\rowcolor{lightgray}
\multicolumn{1}{|c|}{\textbf{Aspect}} &
\multicolumn{1}{c|}{\textbf{\ac{IQ} Samples}} &
\multicolumn{1}{c|}{\textbf{Spectrograms}} \\
\hline
\textbf{Representation} & Raw baseband signal & Derived (time,frequency) representation \\
\hline
\textbf{Information Content} & Amplitude and phase & Time-frequency, energy (but not full phase) \\
\hline
\textbf{Storage/
Processing} & Higher cost & More compact and easier to process \\
\hline
\textbf{Typical Use} & Detailed signal analysis, demodulation, \ac{RF} fingerprinting & Interference analysis, pattern recognition, image-based learning \\
\hline
\end{tabular}
\end{table}

\subsubsection{Reference signals}
Reference signals, also commonly referred to as pilot signals, are predefined signals transmitted by \acp{UE} or \ac{RAN} to support radio procedures such as channel estimation, demodulation, channel-quality assessment, beam management, and link adaptation. Representative reference signals include \ac{CSIRS}, \ac{DMRS}, and \ac{SRS}~\cite{TS38211}. Processing these signals can produce channel-related estimation data, including \ac{CSI}, \ac{CFR}, and \ac{CIR}. Example of using this data in \ac{ORAN}, Anbukani \emph{et al.}~\cite{10571079} study \ac{CSI}-RS processing and transport across \ac{ORAN} fronthaul functional split. Bouknana \emph{et al.}~\cite{11501782} use \ac{UL} \ac{SRS} as the basis for extracting channel information that is exposed from an E2 node to a \ac{NearRTRIC} xApp. Standardized \ac{PHY}-layer indicators derived from these reference signals are discussed separately in Section~\ref{sec:fine_grained_telemetry}.

\subsection{Fine-grained \ac{RAN} telemetry}
\label{sec:fine_grained_telemetry}

\ac{ORAN} architecture exposes telemetry from different components of the \ac{RAN} to support programmable monitoring and data-driven control loops~\cite{10.1145/3570361.3592493}.
Prior work has shown that coarse \ac{ORAN} telemetry may be insufficient for some application (security application for example), motivating the need for finer-grained telemetry~\cite{10.1145/3565474.3569070}.

\begin{figure}[h]
 \centering
 \includegraphics[width=0.9
\columnwidth]{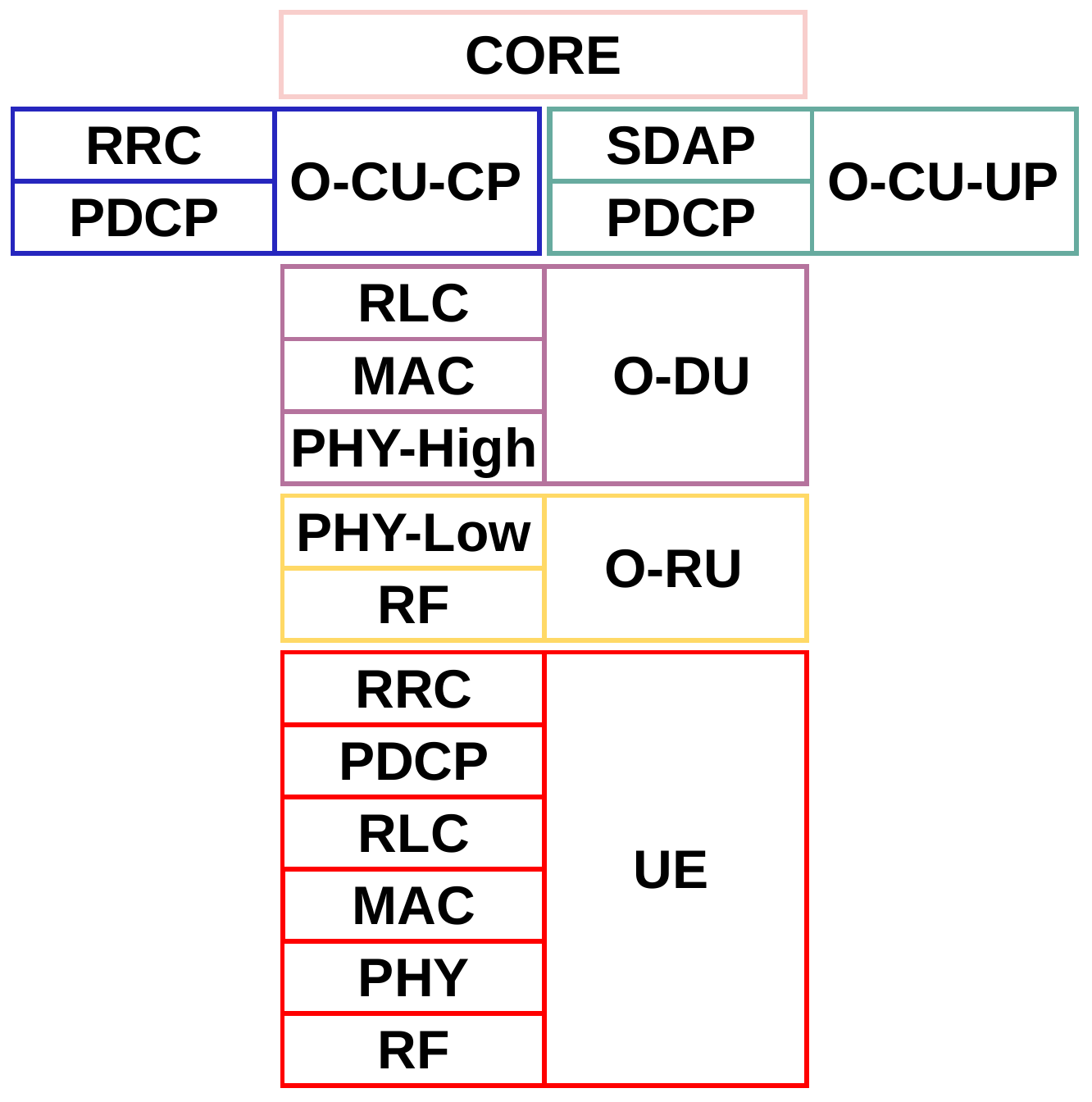}
 \caption{Distribution of protocol stack across \ac{ORAN} components}
 \label{fig:fg_ran_telemetry}
\end{figure}

In disaggregated \ac{ORAN} deployments, control functions are distributed across entities such as the \ac{ODU}, \ac{OCUUP}, and \ac{OCUCP}, while the underlying \ac{ORAN} protocol stack (\ac{NR} stack) spans \ac{PHY}, \ac{MAC}, \ac{RLC}, \ac{PDCP}, \ac{SDAP}, and \ac{RRC} layers~\cite{11028899,10024837,s24031038}. Performance indicators span several layers
of the protocol stack and must be jointly analyzed to understand network behavior~\cite{10.1145/3387514.3405882}. Fig.~\ref{fig:fg_ran_telemetry} illustrates the main protocol layers from which fine-grained \ac{RAN} telemetry can be derived. Since each layer is responsible for different functions, it generates different forms of data. Consequently, fine-grained \ac{RAN} telemetry can be categorized below, according to the protocol layer at which the measurements, metrics or events are produced.

\subsubsection{\ac{PHY} Layer Measurements} represent low-level radio performance indicators. Unlike raw \ac{IQ} samples, which capture unprocessed signal representations, \ac{PHY} measurements correspond to standardized radio metrics computed at the \acp{UE} and the \ac{RAN} components. 

Typical \ac{PHY}-layer measurements and indicators include \ac{RSRP}, \ac{RSRQ}, \ac{SINR}, \ac{BLER}, \ac{CQI}, and \ac{MCS}.

The main \ac{PHY} measurements are standardized by \ac{3GPP}. In particular, \ac{3GPP} \ac{TS} 38.215~\cite{TS38215} defines two main families of \ac{PHY} measurements: (i) SS-based measurements which are derived from \acp{SSB} and are mainly used for signal strength and quality evaluation in procedures such as cell search, initial access, beam selection, and mobility (Examples include SS-\ac{RSRP}, SS-\ac{RSRQ}, and SS-\ac{SINR}); and (ii) \ac{CSI}-based measurements which are derived from \ac{CSIRS} and characterize radio channel conditions for channel-quality reporting, beam management, and link adaptation (Examples include CSI-\ac{RSRP}, CSI-\ac{RSRQ}, and CSI-\ac{SINR}). In addition, \ac{3GPP} \ac{TS} 38.214~\cite{TS38214} specifies CSI reporting quantities such as \ac{CQI}, as well as \ac{PHY} procedures related to modulation and coding, including \ac{MCS}
and transmission performance indicators associated with block error behavior (such as \ac{BLER}).

\subsubsection{\ac{MAC}-layer Metrics} correspond to indicators generated by \ac{MAC} layer of protocol stack. Unlike \ac{PHY}-layer measurements, which describe radio signal conditions, \ac{MAC}-layer metrics reflect how radio resources are allocated and how data buffers are managed during transmission. In \ac{ORAN}, \ac{MAC}-layer metrics are primarily associated with the \ac{ODU}. Typical \ac{MAC} metrics include \ac{PRB} utilization, which represents the amount of radio resources allocated to users or services within a transmission interval; \ac{BSR}, which indicates the amount of data waiting in the \ac{UE} transmission buffers; \ac{HARQ} status, which provides feedback information indicating successful or failed packet transmissions; and scheduling requests and grants, which specify \ac{UL} or downlink resource allocation for data transmission. These metrics are associated with \ac{MAC} procedures defined in \ac{3GPP} \ac{TS} 38.321~\cite{TS38.321}. \ac{MAC} measurement definitions are further specified in \ac{TS} 38.314~\cite{TS38.314}.

Although large number of \ac{PHY}/\ac{MAC} metrics can be collected in \ac{ORAN}, many of them may contain overlapping information. Recent analysis of \ac{PHY}/\ac{MAC} \ac{ORAN} data in~\cite{11372999} showed that several metrics form tightly coupled groups as they are used together when the network decides how to transmit data over the radio link. For example, \ac{RSRP}, \ac{RSRQ}, \ac{SINR}), \ac{CQI}, and \ac{MCS} were grouped together because they are all related to radio channel quality and transmission decisions. These findings suggest that monitoring all metrics simultaneously may introduce redundancy, whereas selecting representative metrics can reduce monitoring complexity.

\subsubsection{\ac{RLC}-layer metrics} is responsible for segmenting and reassembling higher-layer packets into \ac{RLC} \acp{PDU}. Therefore, \ac{RLC} layer metrics describe segmentation, retransmission behavior, and buffer management at \ac{RLC} layer. Typical \ac{RLC} metrics include the RLC retransmission rate, which captures the number of \acp{PDU} retransmitted following transmission failures; \ac{RLC} buffer occupancy, which represents the amount of data waiting in transmission buffers; \ac{RLC} throughput, which measures the rate at which data are successfully delivered through the \ac{RLC} layer; and \ac{RLC} packet-loss indicators. These metrics are associated with \ac{RLC} procedures defined in \ac{3GPP} \ac{TS} 38.322~\cite{TS38.322}.

\subsubsection{\ac{PDCP}-layer Metrics} characterize user-plane packet handling at the \ac{PDCP} layer, which is responsible for functions such as header compression, sequence numbering, duplication handling, reordering, and in-order delivery. In \ac{ORAN} deployments, \ac{PDCP} functionality is typically associated with the \ac{OCUUP}. Therefore, \ac{PDCP} metrics provide visibility into packet delivery behavior above \ac{RLC} layer. Typical \ac{PDCP}-layer metrics include \ac{PDCP} throughput, \ac{PDCP} reordering or reordering-delay related indicators, \ac{PDCP} discard/loss indicators, and \ac{PDCP} packet delay or delivery-delay related indicators. These metrics are associated with \ac{PDCP} procedures defined in \ac{3GPP} \ac{TS} 38.323~\cite{TS38323}. In addition to \ac{RLC}/\ac{PDCP} metrics, user-plane units, such as \ac{RLC} and \ac{PDCP} packets, may also serve as raw inputs for applications such as traffic classification and slice identification~\cite{LACAVA2025111342}.

\subsubsection{\ac{SDAP}-layer Metrics} capture \ac{QoS} of flow-level behavior at the SDAP layer, which is responsible for mapping \ac{QoS} flows to \acp{DRB} and supporting \ac{QoS}-aware treatment in the user plane. In \ac{ORAN} deployments, \ac{SDAP} functionality is typically associated with the \ac{OCUUP}. Therefore, \ac{SDAP} metrics provide visibility into \ac{QoS} differentiation and service-level packet handling above the \ac{PDCP} layer, which is important for \ac{QoS}-aware control and resource management in \ac{ORAN}~\cite{11010844}. Typical \ac{SDAP}-layer metrics include \ac{QoS}-flow to \ac{DRB} mapping statistics, \ac{QoS}-flow throughput, \ac{QoS}-flow packet delay indicators, and \ac{QoS}-flow loss or discard indicators. These metrics are associated with \ac{SDAP} procedures defined in \ac{3GPP} \ac{TS} 37.324~\cite{TS37324}.

\subsubsection{\ac{RRC}-layer Measurements} capture control-plane procedures related to connection establishment, mobility management, measurement reporting, and failure handling. In \ac{ORAN} deployments, \ac{RRC} functionality is typically associated with the \ac{OCUCP}. Therefore, \ac{RRC} measurements provide visibility into \ac{UE} connection state transitions, mobility-related control procedures, and radio-link management above the user-plane layers. Typical \ac{RRC}-layer events and measurements include \ac{RRC} connection setup, reconfiguration, and release events, handover procedures and mobility-related events, \ac{RLF} indicators, and measurement reporting events. These measurements are associated with \ac{RRC} procedures defined in \ac{3GPP} \ac{TS} 38.331~\cite{TS38331}.

\subsection{Aggregated \ac{KPM} data}
Aggregated \acp{KPM} data refer to higher-level performance measurements reported by E2 nodes, such as \ac{ODU}, \ac{OCUCP}, and \ac{OCUUP}, through E2 interface~\cite{O-RAN.WG3.TS.E2GAP,O-RAN.WG3.E2SM-KPM,10439167}. Unlike fine-grained \ac{RAN} telemetry, which captures detailed metrics at specific layers, \acp{KPM} provide more structured and aggregated view of \ac{RAN} performance. In particular,~\textit{\acp{KPM} aggregate metrics from different layers over time windows across monitored entities such as \acp{UE}, cells or slices}~\cite{O-RAN.WG3.E2SM-KPM,ORAN-WG3.E2SM,TS28552,TS28554}. These \acp{KPM} provide valuable insights into the performance of the \ac{RAN} network~\cite{Systemlevel} and help to make decisions to optimize network performance.

Representative \acp{KPM} in \ac{ORAN} cover multiple levels, including\ac{UE}-, RAN-, cell-, slice-, and \ac{QoS}-level measurements~\cite{10102369, O-RAN.WG3.E2SM-KPM}. 

\subsubsection{\ac{UE}-level \acp{KPM}}

\ac{UE}-level \acp{KPM} characterize the performance or state of individual \acp{UE}. Typical examples include \ac{UE} throughput, data volume, delay-related indicators, mobility-related measurements, and connection-related statistics.

\subsubsection{\ac{RAN} Node-level \acp{KPM}}

Node-level \acp{KPM} summarize performance at the reporting-node level, such as the \ac{ODU}, \ac{OCUCP}, or \ac{OCUUP}. Typical examples include aggregate throughput, resource utilization, and other measurements that reflect the operational state of the reporting \ac{RAN} function.

\subsubsection{Cell-level \acp{KPM}}

Cell-level \acp{KPM} summarize performance at the cell level. Typical examples include \ac{PRB} utilization, cell throughput, \ac{CQI}/\ac{MCS}-related summaries, the number of connected \acp{UE}, and cell-level radio resource utilization.

\subsubsection{Slice-level \acp{KPM}}

Slice-level \acp{KPM} aggregate measurements associated with a network slice. Typical examples include slice throughput, slice resource utilization, delay-related indicators, and load-related statistics.

\subsubsection{\ac{QoS}-level \acp{KPM}}

\ac{QoS}-level \acp{KPM} capture performance associated with individual \ac{QoS} flows. Typical examples include \ac{QoS}-flow throughput, packet delay, packet loss or drop indicators, and service-specific delivery behavior.

\ac{KPM} reporting is standardized through the Service Model E2SM-\ac{KPM} from \ac{ORAN} Alliance~\cite{O-RAN.WG3.E2SM-KPM}), which enables the exposure of measurements from \ac{ODU}, \ac{OCUCP}, and \ac{OCUUP}~\cite{ORAN-WG3.E2SM,10439167}. Many of these measurements are aligned with \ac{3GPP} performance-management definitions~\cite{TS28552, TS32425}, while others are extended for \ac{ORAN} reporting needs~\cite{O-RAN.WG3.E2SM-KPM}. It is worth noting that, the units of some \ac{ORAN} \acp{KPM} driven from \ac{3GPP} are replaced by new units such as reporting throughput and data volume in kbit instead of Mbit and in-session activity time in milliseconds instead of seconds~\cite{O-RAN.WG3.E2SM-KPM}.

The \ac{KPM} reporting framework has evolved across \ac{ORAN} specification versions. In Version~1, the \ac{RAN} function ``\ac{KPM} Monitor'' exposed \ac{ODU} cell-related, \ac{OCUCP} cell/\ac{UE}-related, and \ac{OCUUP} bearer-related performance \acp{IE} through periodic \ac{KPM} reports. Version~2 extended this model by advertising the measurements available from the \ac{ODU}, \ac{OCUCP}, and \ac{OCUUP} through the \ac{RAN} Function Definition \acp{IE}, allowing the \ac{NearRTRIC} to subscribe to selected measurements for periodic reporting~\cite{TS104040}.

\subsection{Management-plane and historical data} \label{sec:managmentdata}

Management-plane and historical data correspond to \ac{OAM} data, or in other words to performance, fault and configuration management data (explained in the subsection below), collected mainly from \ac{ODU}, \ac{OCUCP}, and \ac{OCUUP} nodes through O1 interface and handled within \ac{SMO} and \ac{NonRTRIC}. \ac{SMO} collects those management measurements via O1, while rApps at \ac{NonRTRIC} consume these data for \ac{NonRT} control such as policy management, network orchestration or resource optimization~\cite{O-RAN.WG10.TS.OAM,11080542}. Unlike \acp{KPM}, which is designed for \ac{NearRT} control, management-plane data are typically historical, statistical, aggregated on longer time window. Moreover, they are aligned with \ac{3GPP} management specifications (\ac{TS} 28 series)~\cite{O-RAN.WG10.TS.OAM,TS28552,TS103982, 9579445}. Management-plane data are organized into three main categories, namely, \ac{PM}, \ac{FM}, and Configuration Management.

\subsubsection{Performance Management Data}

\ac{PM} data characterize statistical performance data collected from \ac{ORAN} nodes. \ac{PM} data provide visibility into network, resource, and service performance. Typical \ac{PM} data include counters collected over defined periods, including cell, node, and network level measurements, radio-resource utilization, mobility and handover success/failure counters, and reporting metadata such as measurement types and collection timestamps.

\subsubsection{Fault Management Data}

\ac{FM} data characterize alarms, fault notifications, degraded operational states, and abnormal behavior observed in \ac{ORAN} nodes. \ac{FM} data provide visibility into failures, degradation, and abnormal conditions affecting \ac{ORAN} functions and interfaces. Typical \ac{FM} data include alarm identifiers and severity levels, fault timestamps, interface failure notifications, and fault-related logs and reporting metadata.

\subsubsection{ Configuration Management Data}

\ac{CM} data characterize the intended, applied, and observed configuration state of \ac{ORAN} nodes. \ac{CM} data provide visibility into parameter settings and interface configuration. Typical \ac{CM} data include radio configuration parameters, interface and connectivity configuration, software and firmware version information, synchronization parameters, configuration change logs and related metadata. \ac{CM} contains also very valuable topolotical information that be used to recreate the graph of the network.

\subsection{Control-plane and Signaling data}
Control-plane and signaling data include messages and events used to establish and manage \ac{UE} connections, report measurements, and support mobility and handover across \acp{UE} and \ac{RAN} (CU, DU, RU)~\cite{286439,panitsas2024predictive, 8316764}. In 5G networks, \ac{NAS} and \ac{RRC} are the main control-plane signaling protocols. \ac{NAS} facilities signaling messages exchange between \ac{UE} and 5G Core~\cite{TS24501}, while \ac{RRC} carries signalling messages and data-plane parameters~\cite{286439} to support connection setup, reconfiguration, measurement reporting, and handover-related behavior between \ac{UE} and \ac{gNB}~\cite{TS38331}. On the other hand, in \ac{ORAN}, this type of data are associated with \ac{UE}, \ac{OCUCP} (the logic node hosting \ac{RRC} protocol), and \ac{ODU}, and includes \ac{RRC} signaling, mobility and handover events and random access procedures. 

 \begin{figure*}[h]
\centering
\includegraphics[width=0.5\textwidth]{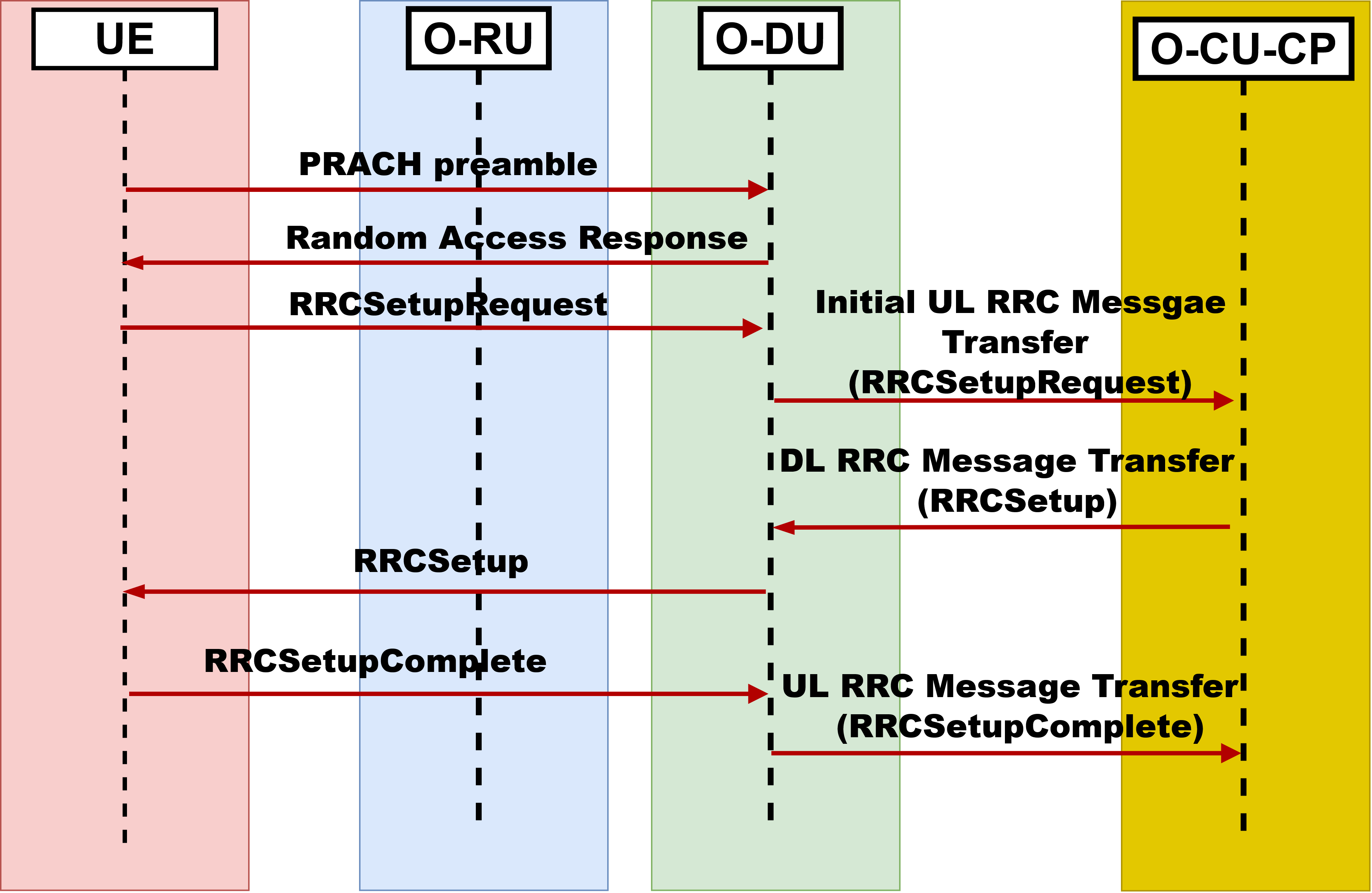}
\caption{Control-plane signaling flow in \ac{ORAN} during \ac{UE} \ac{RRC} connection establishment}
\label{fig:oran_cp_flw}
\end{figure*}

To explain an example of signalling messages, Fig.~\ref{fig:oran_cp_flw} shows, in a simple way, the control-plane exchange messages during \ac{RRC} connection establishment of \ac{UE} in \ac{ORAN}. In this architecture, base station (\ac{gNB}) functions are split between \ac{ORU} and \ac{ODU} through \ac{OFH} interface (split 7.2 is considered)~\cite{TS103982}, and between the \ac{ODU} and \ac{OCUCP} through the F1-C interface, which uses \ac{F1AP} for control signalling exchange~\cite{TS38473}. First, \ac{UE} initiates random access by transmitting a \textit{\ac{PRACH}preamble}, to which the network component \ac{ODU} responds with a \textit{Random Access Response}providing the \ac{UL} grant~\cite{TS38.321, 7945855}. In the disaggregated architecture of \ac{ORAN}, these first two messages are handled at \ac{ORU}/\ac{ODU} side, whereas the remaining initial-attachment signaling is processed at \ac{OCUCP}~\cite{TS103982}. After \textit{Random Access Response} provides the \ac{UL} grant, \ac{UE} sends an \textit{RRCSetupRequest}. The \ac{ODU} forwards this message to the \ac{OCUCP} over F1-C using the \textit{Initial \ac{UL} \ac{RRC} Message Transfer}, the \ac{OCUCP} then generates \textit{RRCSetup} and returns it via the \textit{downlink \ac{RRC} Message Transfer}. Finally, the \ac{UE} completes the procedure with \textit{RRCSetupComplete}, which the \ac{ODU} relays to the \ac{OCUCP} using the \textit{\ac{UL} \ac{RRC} Message Transfer} and which may also carry a dedicated \ac{NAS} message to initiate registration toward the core network~\cite{TS38.321,TS38331,TS38473}.

Many of these signaling messages can be exploited by \ac{ORAN} applications for monitoring, protection, and control. For example, in~\cite{10226043}, copies of \textit{Random Access Response} and \textit{RRCSetupRequest} messages are delivered to an xApp, which uses parameters such as \ac{TA} and the long-term statistics of \ac{RRC} setup requests to detect signaling-storm attacks at an early stage. In contrast,Chatzimiltis \emph{et al.}~\cite{chatzimiltis2025surviving} treats the \ac{UE} initial signaling as an input to an \ac{ORAN} application for congestion control and resource adaptation. Moreover, recent studies in \ac{ORAN} also show that~\textit{mobility-related control data} can also be exploited by xApps, where \ac{UE} measurements and handover events are collected through E2 interface and used to support predictive handover control and other mobility optimization applications~\cite{panitsas2024predictive}. However, unlike aggregated \ac{KPM} telemetry, control-plane data are sparse, and highly sensitive, since its exposure may reveal \ac{UE} identity, location, or other private data; therefore, secure handling of such signaling information is important in \ac{ORAN}~\cite{liyanage_open_2023}. We will discuss the privacy challenge in Section~\ref{sec:Challenges}.

\subsection{Cross-domain data} 
Cross-domain data refer to data that are imported from multiple domains connected to \ac{ORAN} network. This includes \ac{EI}, policies and \ac{AIML} management data from \ac{SMO}, service and session context from the core network, and traffic observations from monitoring systems.

\subsubsection{Enrichment Information}

Through A1 interface, \ac{NonRTRIC} can provide additional information called \ac{EI} to the \ac{NearRTRIC}~\cite{TS103983,10329915, s21248173, 9579445}. \ac{EI} is not available to \ac{RAN} itself, instead it represents a supplementary context provided in addition to what is already available inside \ac{RAN}. \ac{EI} may be derived from internal or external sources~\cite{TS103983}. \ac{EI} is supported in \ac{ORAN} by the \ac{A1EI} service. The operations and \acp{API} of this service are specified by \ac{A1AP}~\cite{TS103987}. \ac{EI} types are defined in the \ac{A1TD} specification~\cite{TS103988}. The standard distinguishes between two kinds of \acp{EI}: (i) \ac{ORAN}-defined \ac{EI} types; and (ii) \ac{EI} types defined by another entity. In the current version of the standard, the explicitly defined \ac{ORAN} \ac{EI} type is \textit{\ac{UE} location and velocity information}. On the other hand, \ac{EI} types defined by another entity are not explicitly listed in the standard; This is left open for operators or vendors depending on \ac{ORAN} deployment needs. Possible examples of \acp{EI} can be \ac{SMO} level observations, service metadata, external inputs, or \ac{AIML}-related data used for training and inference~\cite{10329915}. Topological information can also be provided by \ac{EI}.

\subsubsection{Policies}
Policies in \ac{ORAN} represent high-level control directives that guide the behavior of \ac{NearRTRIC} applications and, indirectly, E2 nodes. Unlike telemetry or enrichment data, which describe the network state, policies are \textit{prescriptive}: they define constraints, objectives, or operational rules that control decisions under specific network conditions. \ac{NonRTRIC} applications translate operator intent and optimization goals into policy objects. These policies are conveyed to \ac{NearRTRIC} via the A1 interface using the A1 Policy Management Service (A1-P)~\cite{TS103987}. Policies do not directly carry radio control commands. Instead, they constrain the decision space of xApps. Examples of \ac{ORAN} policies can be categorized as follows:

\begin{itemize}

 \item Resource Management Policies: define communication and computation resources allocating limits. For example, Sroka \emph{et al.}~\cite{10529320} present Policy-driven traffic steering. Moreover, policy-aware fair resource sharing is studied in~\cite{10.1145/3639043}.

 \item \ac{QoS} Policies: specify performance objectives such as latency, throughput, or reliability. Marinova and Leon-Garcia~\cite{10439167} discuss \ac{SLA} objectives at \ac{NonRTRIC} and its influence on \ac{NearRT} xApps. In addition, traffic steering under \textit{constraints} defined by operators is illustrated in~\cite{10529320}.

 \item Mobility Policies: regulate handover behavior, cell reselection strategies, and traffic distribution across cell/slices. Traffic steering is demonstrated in~\cite{11317189}, where \ac{NonRT} policies constrain \ac{NearRT} decisions. Mobility-aware (policy) load balancing is further examined in~\cite{10529320}.

 \item Energy Policies: define dynamic power-saving modes and energy-aware optimization objectives. 
 The white paper~\cite{ORAN_Energy_Savings_Whitepaper} describes energy-saving control mechanisms that can be implemented through policy-based control.

 \item Security Policies: impose constraints related to anomaly thresholds, or mitigation strategies in response to detected threats. Marinova and Leon-Garcia in~\cite{10439167} discuss control frameworks, where security policies are translated into xApps constraints.

\end{itemize}

\subsubsection{\ac{AIML} Model Management Data}

These refer to the data exchanged to support the training, monitoring, and operational use of \ac{AIML} models in \ac{ORAN}. Through the A1 interface, the \ac{SMO}/\ac{NonRTRIC} and \ac{NearRTRIC} can exchange \ac{AIML} model training-related information using \ac{A1ML} service~\cite{TS103983, O-RAN.WG2.TS.A1GAP}. A1 specifications define procedures for querying \ac{AIML} model training capabilities, requesting model training, cancelling training, and querying training. Moreover, this category includes model descriptors, training metadata, model-selection information, and performance feedback~\cite{10604823}.

\subsubsection{Core network data}
Core network data consist of the control and user plane data exchanged between network core and \ac{ORAN} component (\ac{OCU}) via \ac{3GPP} NG interfaces. The control plane (\ac{OCUCP}) via N2 interface connects to the \ac{AMF} of the core, while the user plane (\ac{OCUUP}), via N3 interface connects to the \ac{UPF}~\cite{TS103982, 10711198}. For delivering services in the mobile networks, the core provides essential subscriber, session, service and user context metadata that link \ac{RAN} with the core network. Such metadata (explained below) can be used by the application of \ac{ORAN}.

\textit{Subscriber context}: establishes the subscriber’s identity within the network. It answers the question: Who is the subscriber? It includes identity data such as \ac{SUPI}, authentication data, and subscription data~\cite{TS23003}. Those are used by the core to authenticate \acp{UE} and determine services/slices the subscriber is authorized to access.

\textit{Session context}: centers on the management of \ac{PDU} sessions, which represent the logical connection between \acp{UE} and the network~\cite{10925450}. It answers the question: What session connection is currently open for this user? It includes session state, associated attributes, and tunnel data. This is essential for mobility management applications, as it ensures that the connection state remains synchronized as a user moves between different zones. 

\textit{Service context}: bridges application requirements with radio resource allocation. It answers the question: What \ac{QoS} does the traffic need and how should this traffic be treated? These data capture information related to \acp{QFI}, and service requirements. In \ac{ORAN}, the core network determines the \ac{QoS} treatment of traffic and maps \ac{IP} flows to QoS flows, while \ac{SDAP} function in \ac{OCUUP} maps \ac{QoS} flows to \acp{DRB}, ensuring that traffic receives the appropriate latency, throughput, and reliability according to the service requirements~\cite{TS24501, s23021047, 10.1145/3570361.3615737}.

\textit{User-plane tunnel context}: it is the transport-level metadata used for data delivery between the Access Networks and \ac{UPF}. It answers Which packet tunnel should carry the data? It centers on the \ac{GTPU}~\cite{TS29281} and involves \acp{TEID}. This context is vital for routing user-plane traffic between \ac{UPF} in the core and the \ac{OCUUP}, managing the encapsulation headers that allow packets to traverse the backhaul network~\cite{10.1145/3719027.3765206}.

Besides the above categories of metadata that the core network provides to network applications, the actual traffic carried over the user plane can also be observed and analyzed; this is discussed separately in the next section.

\subsubsection{Network Traffic Data}
Network traffic data refer to the actual user-plane traffic observed in \ac{ORAN} components especially at monitoring points in the core network (\ac{UPF}), or functions that can directly access user-plane data at \ac{RAN}. Such data can be analyzed at different granularities, including packet, flow, connection, and host level~\cite{8428412}.
 
\textit{Packet-level traffic data}: These data correspond to individual packets and includes packet headers, timestamps, packet sizes, and payload (when there are no concerns about privacy). The packet level provides the most detailed view of network activity; however, it also introduces the highest collection overhead and privacy sensitivity~\cite{8428412}. In \ac{ORAN}, packet-level inspection is performed at \ac{UPF} or inside the \ac{RAN} components. For example, Tsourdinis \emph{et al.}~\cite{10.1145/3636534.3697311} analyze \ac{UPF} traffic to support intrusion detection application and resource allocation decisions in \ac{ORAN}. Typical packet-level traffic data include source and destination \ac{IP} addresses, source and destination port numbers, protocol type, packet timestamps or inter-arrival timing information, packet size and byte counts, packet payload, and \ac{TCP} flags.

\begin{figure*}[h]
 \centering
 \includegraphics[width=0.7\textwidth]{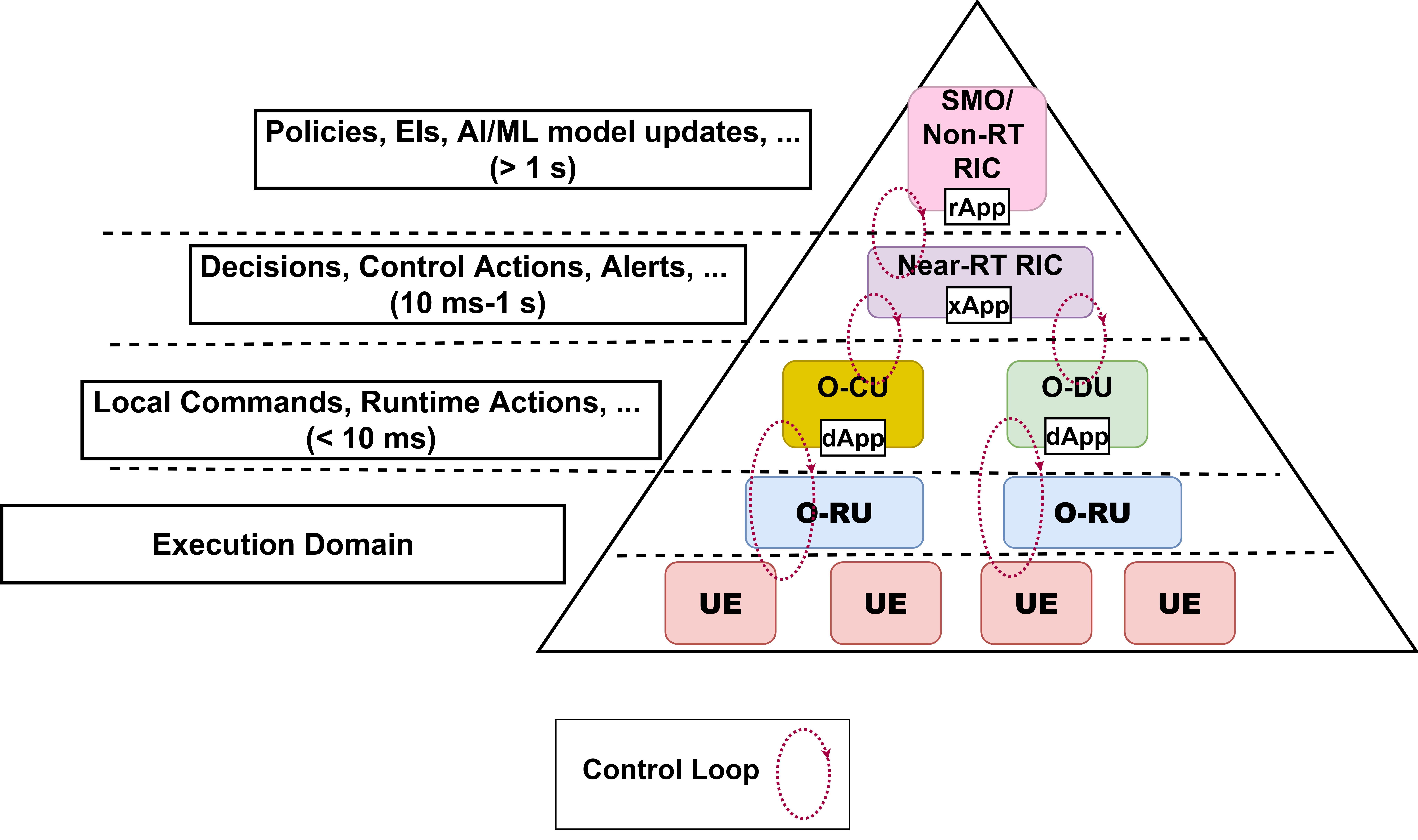}
 \caption{Hierarchical view of intelligence-derived data and control loops in \ac{ORAN}}
 \label{fig:intelligence_hierarchy}
\end{figure*}

\textit{Flow-level traffic data}: These data aggregate packets that share common attributes, typically identified by the 5-tuple attributes: (i) source \ac{IP} address, (ii) destination \ac{IP} address, (iii) source port, (iv) destination port, and (v) transport protocol. Compared with packet-level data, flow-level traffic provides a more compact representation of user activity, making it suitable for monitoring and traffic classification in high-speed networks. In \ac{ORAN}, flow-level traffic is usually observed from the core network and the transport-network, where conventional flow-monitoring mechanisms such as NetFlow and sFlow may be used. However, for \ac{RT} \ac{ORAN} scenarios, a prior work notes that these packet-sampling mechanisms may be insufficiently fine-grained or timely, motivating the use of more advanced telemetry approaches such as \ac{INT} and \ac{IOAM}~\cite{11010844}. For example, Xavier \emph{et al.}~\cite{xavier2024crossdomainaiearlyattack} use the flow level data for early detection of malicious flows in \ac{ORAN}. On the other hand, Groen \emph{et al.}~\cite{10622798} show that traditional 5-tuple flow information is generally not directly available to \ac{NearRTRIC}, which is why \ac{ORAN} applications such as traffic classification often relies on \ac{KPM} instead. Typical flow-level traffic data include source and destination \ac{IP} addresses, source and destination port numbers, transport protocol, flow duration, flow direction, number of packets per flow, number of bytes per flow, flow start and end times, and inter-packet timing statistics.

\textit{Connection-level traffic data}: These data provide a coarser-grained view than packet/flow-level data by describing the communication state between endpoints over a period of time. It shows what communication connection is currently active for a user, and how it behaves. In \ac{ORAN}, such data are usually derived from aggregate user-plane observations collected from the transport/core networks. This level is useful for traffic anomaly detection, since persistent, short-lived, unbalanced, or abnormal connections may indicate malicious activity~\cite{10.1145/3636534.3697311}. Typical connection-level traffic data includes connection count/duration/state/type, total number of packets exchanged per connection, total number of bytes exchanged per connection, and connection start and end times.

\textit{Host-level traffic data}: These data refer to observations collected from the host or software platform that runs a network function. It typically includes \ac{CPU} and memory utilization, process-level statistics, and others. In \ac{ORAN}, such data are especially useful for applications deployed close to network functions, such as dApps running in E2 nodes (\ac{ODU} or \ac{OCU})~\cite{crespo2025energy}. Typical host-level traffic data include \ac{CPU}/memory usage, process/thread statistics, cache indicators, local logs and runtime events, buffering indicators, and host-side timing and performance statistics.

\begin{table*}[h]
\centering
\renewcommand{\arraystretch}{1.35}
\caption{\ac{ORAN} Data Classes, Sources, Interfaces, and Consumers}
\label{tab:oran_data_classification}
\begin{tabular}{|p{2.7cm}|p{2.9cm}|p{4.5cm}|p{3.4cm}|p{2.5cm}|}
\hline
\rowcolor{lightgray}
\multicolumn{1}{|c|}{\textbf{Data Class}} &
\multicolumn{1}{c|}{\textbf{Source}} &
\multicolumn{1}{c|}{\textbf{Examples}} &
\multicolumn{1}{c|}{\textbf{Access Interfaces}} &
\multicolumn{1}{c|}{\textbf{Consumers}} \\
\hline

Radio-level data &
\ac{UE}, \ac{ORU} &
\ac{IQ} samples, Spectrograms, low-level signal representations &
Uu; \ac{OFH} U-plane &
dApps, xApps
\tabularnewline
\hline

Fine-grained \ac{RAN} telemetry &
\ac{ODU}, \ac{OCUCP}, \ac{OCUUP} &
\ac{PHY}/\ac{MAC}/\ac{RLC}/\ac{PDCP}/\ac{SDAP}/\ac{RRC}-layer metrics &
E2, E3 &
xApps, dApps
\tabularnewline
\hline

Aggregated \ac{KPM} data &
\ac{ODU}, \ac{OCUCP}, \ac{OCUUP} &
\ac{UE}/cell/ \ac{RAN} node/slice/\ac{QoS} level \acp{KPM} &
E2 &
xApps, rApps
\tabularnewline
\hline
Management-plane and historical data &
\ac{ORU}, \ac{ODU}, \ac{OCUCP}, \ac{OCUUP}, \ac{NearRT} RI &
\ac{PM}/\ac{FM}/\ac{CM} data &
O1 &
rApps
\tabularnewline
\hline

Control-plane and signaling data &
\ac{UE}, \ac{ODU}, \ac{OCUCP}, core &
\ac{RRC} signaling, random-access events, mobility/handover events, \ac{NAS} messages &
Uu, F1-C, Xn-C, NG-C, E2 &
xApps, rApps
\tabularnewline
\hline

Cross-domain data &
Core network, \ac{SMO}, external systems &
\ac{EI}, policies, \ac{AIML} management data, core network data, Network traffic data &
A1, O1, N2/N3, external \acp{API}&
rApps, xApps, dApps
\tabularnewline
\hline

Intelligence-derived data &
xApps, rApps, dApps&
Policies, Control Decisions, recommendations &
E2, A1, O1, R1, local control loops, internal \ac{RIC} messaging &
E2 nodes, xApps, rApps, dApps
\tabularnewline
\hline

User intent data &
Operators, users &
Natural-language intents, structured intents, \ac{JSON}/\ac{YAML} intent templates &
R1, northbound \acp{API}, external \acp{API}, agent interfaces &
xApps, rApps, dApps, \ac{SMO},
\tabularnewline
\hline
\end{tabular}
\end{table*}

\subsection{Intelligence-derived data}

These data correspond to outputs produced by \ac{ORAN} applications after processing measurements data, historical records, or local observations. These outputs may take the form of control decisions, policy descriptors, predictions, anomaly scores, classification labels, or other derived recommendations that are then consumed by other \ac{ORAN} applications. Depending on where the application is deployed, such outputs can be conveyed through E2, A1, local \acp{API}, or internal \ac{RIC} messaging frameworks~\cite{11010844}. In particular, xApps typically produce \ac{NearRT} control decisions for enforcement at E2 nodes, rApps mainly produce high-level policies and long-timescale guidance for xApps, and dApps produce \ac{RT} local action outputs close to \ac{ODU}/\ac{OCU} nodes~\cite{9846950}. Figure~\ref{fig:intelligence_hierarchy} illustrates this hierarchy of intelligent control in \ac{ORAN}. At the top, the \ac{SMO}/\ac{NonRTRIC} layer hosts rApps, which generate long-timescale policies, \ac{EI}, and \ac{AIML} model guidance. At the middle layer, the \ac{NearRTRIC} hosts xApps, which transform telemetry into \ac{NearRT} decisions delivered to E2 nodes. At the lower layer, dApps are deployed directly on \ac{OCU}/\ac{ODU} nodes to generate immediate local actions using lower-layer, host-level, or user-plane-adjacent data. The figure also highlights the closed-loop nature of these intelligence flows.

\subsubsection{xApp output data}refer to the intelligence generated by xApps after processing \ac{ORAN} data. Such outputs may include control decisions or other actions intended to influence \ac{RAN} behavior. In many cases, these outputs are conveyed to the \ac{ODU} or \ac{OCU} over E2 interface, where \ac{E2AP}~\cite{TS104039} provides message-exchange procedures, and the corresponding \ac{E2SM} defines the semantics of the control action. In particular, xApps commonly rely on the \ac{E2SMRC}~\cite{TS104040} service model to deliver enforcement actions to E2 nodes~\cite{10.1145/3636534.3697325}. Examples of xApp outputs include control decisions for \ac{RAN} parameter adjustment, resource management commands, traffic steering actions, anomaly scores, alerts, or risk indicators, thresholds, and messages exchanged with other \ac{RIC} functions or xApps.

\subsubsection{rApp output data} refer to the intelligence generated by rApps after processing \ac{ORAN} data. Such outputs may include policies, optimization recommendations, predicted traffic or load trends, configuration guidance, model-related updates, or \ac{EI} intended to guide control loops (in dApps and xApps). In many cases, these outputs are conveyed toward the \ac{NearRTRIC} through A1 interface, where they are used to constrain the behavior of xApps~\cite{giannopoulos2026interoperable,10225883}.

\subsubsection{dApp output data} refer to the intelligence generated by applications deployed directly on \ac{ORAN} nodes (\ac{ODU}, \ac{OCU}), where host data may be directly accessible. Such outputs may include local control actions or aggregated local state intended for coordination with higher layer controllers. Unlike xApps and rApps, dApp outputs are typically applied through local \acp{API} or host-level control mechanisms, rather than through standard E2 or A1 interfaces~\cite{olimpieri2025libiq,crespo2025energy}. Examples of dApp output include local \ac{CPU}/thread commands, frequency control, or \ac{RT} anomaly traffic detection.

\subsection{User Intent data}

User intent data refer to objectives expressed by operators, users, or applications to guide \ac{ORAN} behavior. Intent data describe the desired outcome of network operation. In traditional \ac{ORAN} applications, such objectives are usually encoded manually as policies, thresholds, or configuration parameters. However, in Agentic \ac{AI}, intent becomes an explicit data type consumed by \ac{LLM}-based agents. These agents interpret natural-language or structured intents, reason over the current network state, and translate them into machine-readable policies, or control actions~\cite{bimo2025intent,osti_10628444,elkael2025agentran}. User intent data may appear in two main forms: (i) \textit{natural-language intent}: text instructions from users or operators, such as "\textit{reduce energy consumption for this slice}"; and (ii) \textit{structured intent}: template-based representations of the requested goal, typically encoded using formats such as \ac{JSON}. From a data-centric perspective, user intent is important because it links human requirements with \ac{ORAN} control loops. In Agentic \ac{AI} frameworks, an intent is commonly processed through a pipeline: the agent parses the input, extracts objective and constraints, checks consistency and safety, maps the result to control actions, and then monitors feedback from the network.

Table~\ref{tab:oran_data_classification} summarizes the classification of these data classes (mentioned in the Section~\ref{sec:SourceDomain}) in terms of source, representative examples, interfaces, and consumers (applications).

\section{Accessibility View of \ac{ORAN} Data}
\label{sec:Accessibility}
This section discusses how \ac{ORAN} data become accessible through three categories of access mechanisms: native \ac{ORAN} interfaces, \ac{3GPP} interfaces, and emerging Agentic \ac{AI} interoperability \acp{API}. Table~\ref{tab:accessibility_interfaces} summarizes these interfaces, including the connected components, accessible data, and consumers.

\begin{table*}[!h]
\centering
\renewcommand{\arraystretch}{1.6}
\caption{\ac{ORAN} Data Accessibility Interfaces}
\label{tab:accessibility_interfaces}

\begin{tabular}{|p{1.7cm}|p{1.2cm}|p{4cm}|p{6cm}|p{3cm}|}
\hline
\rowcolor{lightgray}
\multicolumn{1}{|c|}{\textbf{Category}} &
\multicolumn{1}{c|}{\textbf{Interface}} &
\multicolumn{1}{c|}{\textbf{Connected Components}} &
\multicolumn{1}{c|}{\textbf{Accessible Data}} &
\multicolumn{1}{c|}{\textbf{Consumers}} \\
\hline

\multirow{6}{*}{\ac{ORAN}}
& E2
& \ac{NearRTRIC} $\leftrightarrow$ E2 nodes
& \acp{KPM}, fine-grained \ac{RAN} telemetry, \ac{RAN} control messages
& xApps \\
\cline{2-5}

& A1
& \ac{NonRTRIC} $\leftrightarrow$ \ac{NearRTRIC}
& Policies, \ac{EI}
& rApps, xApps \\
\cline{2-5}

& O1
& \ac{SMO} $\leftrightarrow$ \ac{ORAN} functions
&\ac{PM}, \ac{FM}, \ac{CM} 
& rApps \\
\cline{2-5}

& O2
& \ac{SMO} $\leftrightarrow$ O-Cloud
& Monitoring, alarms
& \ac{SMO}/rApps \\
\cline{2-5}

& R1
& rApps $\leftrightarrow$ \ac{NonRTRIC}
& \ac{AIML} workflow data
& rApps \\
\cline{2-5}

& E3
& dApps $\leftrightarrow$ \ac{ODU}/\ac{OCU}
& \ac{IQ} samples, \ac{PHY} data, local runtime telemetry, dApp control messages
& dApps \\
\hline

\multirow{7}{*}{\ac{3GPP}}
& Uu
& \ac{UE} $\leftrightarrow$ \ac{ORU}
& Radio signals, \ac{UE} measurements, user traffic
& \ac{UE}/\ac{RAN} \\
\cline{2-5}

& F1-C
& \ac{ODU} $\leftrightarrow$ \ac{OCUCP}
& F1-C: control-plane signalling
& Signalling\\
\cline{2-5}

& F1-U
& \ac{ODU} $\leftrightarrow$ \ac{OCUUP}
& F1-U: user-plane traffic and bearer data
& User-plane forwarding \\
\cline{2-5}

& E1
& \ac{OCUCP} $\leftrightarrow$ \ac{OCUUP}
& \ac{UE} context, bearer/session coordination
& \ac{OCU} functions \\
\cline{2-5}

& NG-U/N3
& \ac{OCUUP} $\leftrightarrow$ 5G Core (\ac{UPF})
& User-plane tunnel and packet/flow data
& Core/\ac{RAN} \\
\cline{2-5}

& NG-C/N2
& \ac{OCUCP} $\leftrightarrow$ 5G Core (\ac{AMF})
& Core control and session context
& Core/\ac{RAN} \\
\cline{2-5}

& Xn/X2
& Inter-node \ac{RAN} interfaces
& Mobility, handover, inter-node control/user-plane data
& \ac{RAN} nodes \\
\hline

\multirow{4}{*}{Agentic \ac{AI} \ac{API}}
& \acs{MCP}
& \ac{LLM} agents $\leftrightarrow$ tools/data sources
& Resources, tools, logs, \ac{KPM} streams, configuration files, operational context
& Agentic \ac{AI} \\
\cline{2-5}

& \acs{ACP}
& Heterogeneous Agents
& task metadata, session-aware context
& \ac{AI} agent coordination \\
\cline{2-5}

& \acs{A2A}
& Client agent $\leftrightarrow$ Remote agents
& Agent tasks and messages
& \ac{AI} agents collaboration \\
\cline{2-5}

& \acs{ANP}
& Peer Agents
& Agent identity, discovery, authentication, peer-to-peer coordination data
& Federated agentic systems \\
\hline

\end{tabular}
\end{table*}

\subsection{\ac{ORAN} Interfaces }
Native \ac{ORAN} interfaces include:

\subsubsection{E2 Interface}
The E2 interface connects \ac{NearRTRIC}to E2 nodes (\ac{ODU}, \ac{OCU}). It is the main interface through which \ac{RAN} telemetry is exposed and through which near-RT control actions are back to RAN~\cite{O-RAN.WG3.TS.E2GAP}. The main data directly associated with E2 are fine-grained \ac{RAN} telemetry, \ac{KPM}, and intelligent control outputs sent from \ac{NearRTRIC} to E2 nodes. An E2 message consists of two main protocol components: (i) \ac{E2AP}, which defines signalling procedures and common \acp{IE} of the interface, and (ii) \ac{E2SM}, which defines the structure and semantics of services, such as \ac{KPM} reporting or \ac{RAN} control~\cite{O-RAN.WG3.E2SM-KPM,10329918,TS104039,TS104040}. Typical \ac{E2AP} procedures include support-function procedures such as E2 Setup, E2 Node Configuration Update, E2 Connection Update, and E2 Removal, as well as \ac{RIC} service procedures such as \ac{RIC} Subscription, \ac{RIC} Indication, and \ac{RIC} Control~\cite{TS104039,TS104040}. E2 messages are carried over \ac{SCTP}~\cite{RFC9260} and \ac{IP}, meaning that an \ac{E2SM} payload is encapsulated inside an \ac{E2AP} message, which is then transported over \ac{SCTP}/\ac{IP}. E2 supports several services that define how \ac{NearRTRIC} and E2 nodes exchange data or control actions:

\begin{itemize}
\item{E2 Report} is used when \ac{NearRTRIC} requests measurements or status information from E2 nodes. After a subscription is established, E2 node can send \textit{RIC Indication} messages either periodically or when predefined conditions are met. This service is mainly used to expose \acp{KPM} from \ac{RAN} toward xApps~\cite{TS104039,TS104040,O-RAN.WG3.E2SM-KPM}.

\item{E2 Insert} is used to notify \ac{NearRTRIC} when a specific internal condition or radio-resource-management event occurs and may require \ac{RIC} involvement. In this case, E2 node sends a \textit{\ac{RIC} Indication} message associated with that event and may temporarily pause the related local procedure while waiting for \ac{RIC} response. Insert is used when E2 node needs to inform \ac{RIC} about an event that may affect immediate decision making~\cite{TS104039,TS104040}.

\item{E2 Control} is used by \ac{NearRTRIC} to send actions to E2 nodes. This is realized through a \textit{RIC Control Request}, followed by an acknowledgment or failure response from the node. Through this service, xApps can influence exposed \ac{RAN} functions, parameters, or behaviours at the \ac{ODU}, \ac{OCUCP}, or \ac{OCUUP}~\cite{TS104039,TS104040,10.1145/3636534.3697325}.

 \item{E2 Policy} is used to install operating rules at E2 nodes. Rather than sending a direct command every time an event occurs, the \ac{RIC} provides a policy that specifies how E2 node should behave under certain conditions~\cite{TS104040}. In this way, the decision logic can remain at the node side while still being guided by \ac{NearRTRIC}.

\item{E2 Query} is used when \ac{NearRTRIC} needs to request specific information from E2 node on demand. Unlike report, which is subscription based, query is initiated when \ac{RIC} wants a particular piece of information. This makes it useful for state inspection, when periodic reporting is not sufficient~\cite{TS104040}.

\end{itemize}

\begin{figure}[h!]
\centering
\includegraphics[width=0.5\textwidth]{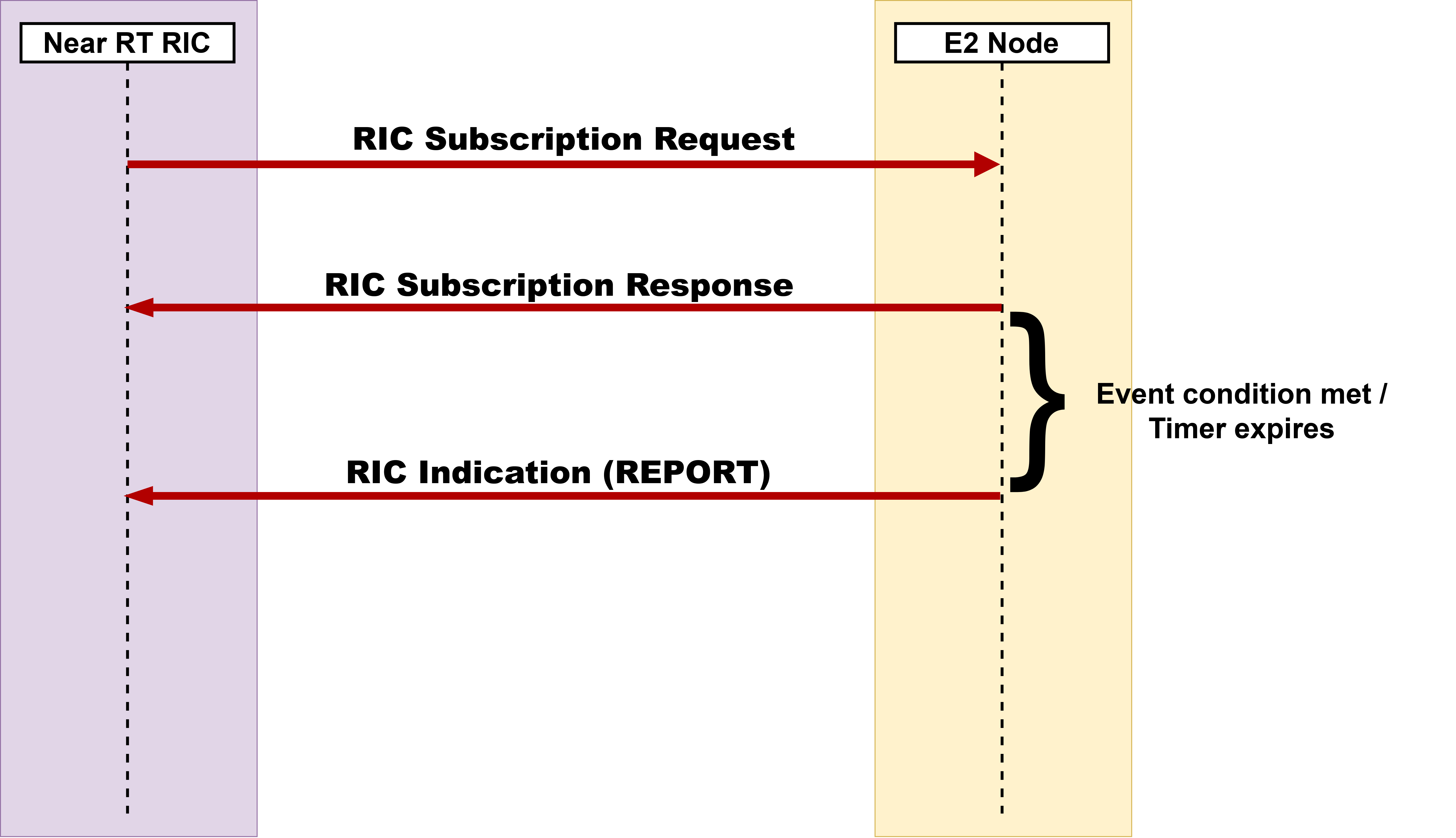} 
\caption{Telemetry workflow over E2 interface us the REPORT service}
\label{fig:e2_telemetry_workflow}
\end{figure}

The actual service data are carried in \ac{E2AP} \acp{IE} whose payload semantics are defined by the selected \ac{E2SM}. For example, in a telemetry workflow using the REPORT service (shown in Fig.~\ref{fig:e2_telemetry_workflow}), the \ac{NearRTRIC} initiates a \textit{RIC Subscription Request} toward E2 node. This subscription establishes a \ac{RIC} subscription for a target \ac{RAN} function, consisting of an event trigger and a sequence of \ac{RIC} Service Actions. If the subscription is accepted, the E2 node responds with a \textit{RIC Subscription Response}. Subsequently, upon occurrence of the configured \ac{RIC} Event Trigger condition, the E2 node sends \textit{RIC Indication} messages to the \ac{NearRTRIC}. These messages include a \textit{RIC Indication Header} and a \textit{RIC Indication Message}, whose detailed content is defined by the selected service model, for example \ac{E2SMKPM} for performance measurements~\cite{TS104039,TS104040}.

\begin{figure}[h!]
\centering
\includegraphics[width=0.44 \textwidth]{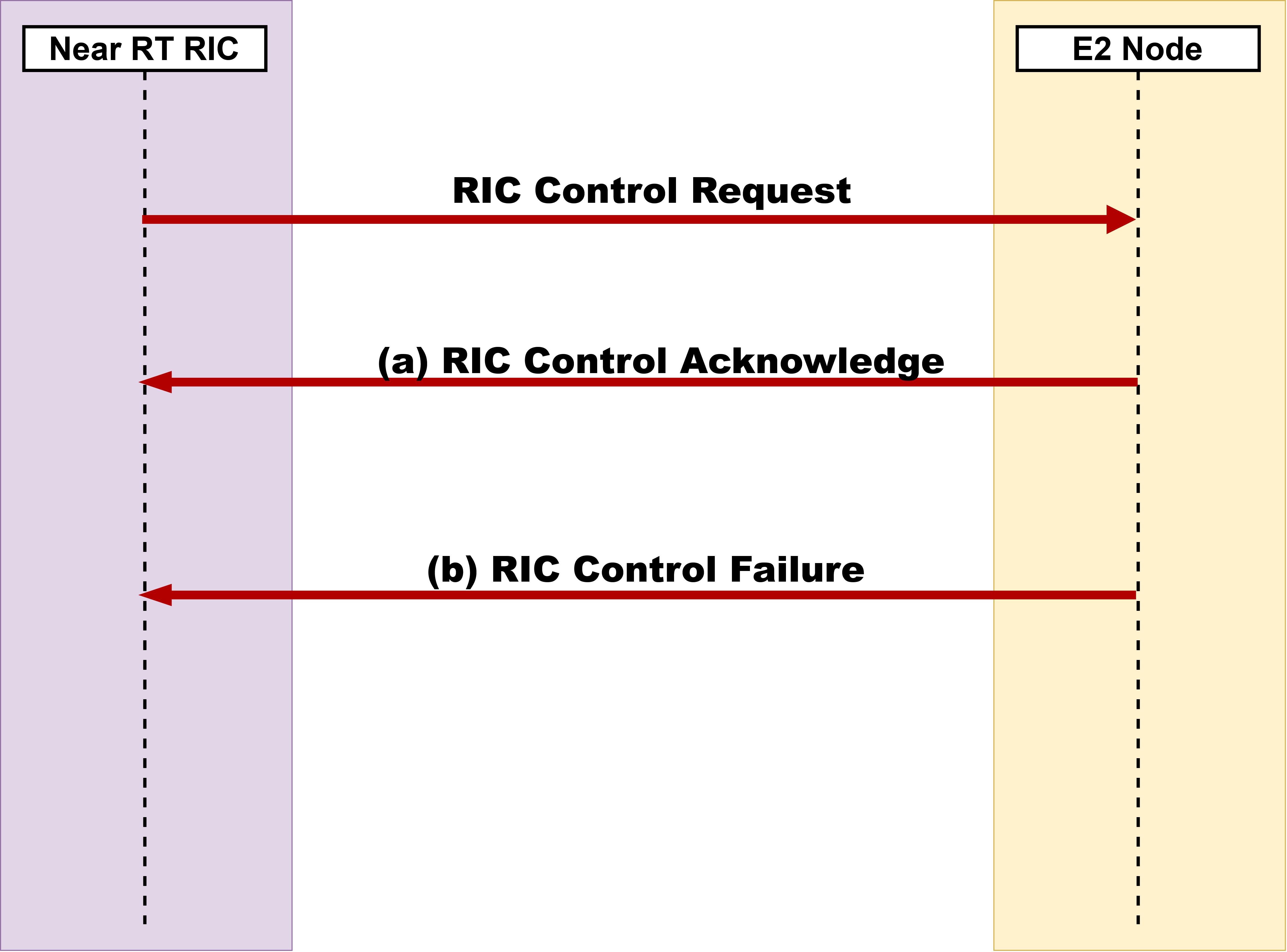}
\caption{Control workflow over E2 interface using the CONTROL service} 
\label{fig:e2_control_workflow}
\end{figure}

Similarly, in a control workflow using the CONTROL service (shown in Fig.~\ref{fig:e2_control_workflow}), the \ac{NearRTRIC} initiates a \textit{RIC Control Request} toward E2 node. This message carries service-model-specific control content, including a \textit{Control Header} and a \textit{Control Message}, whose detailed structure is defined by the selected control-oriented service model, for example \ac{E2SMRC}. If the requested action is successfully processed, the E2 node returns a \textit{RIC Control Acknowledge} when the optional \textit{RIC Control Ack Request} \acp{IE} is set to \textit{Ack} or is not present; if that \acp{IE} is set to \textit{NoAck}, no acknowledge is sent. If the requested action cannot be executed, or if the request contains an invalid or unsupported control specification, the E2 node returns a \textit{RIC Control Failure}. Therefore, \ac{E2AP} defines the signalling procedures and protocol envelope through which the entities communicate, whereas \ac{E2SM} defines the structure and semantics of the control data carried inside those procedures~\cite{O-RAN.WG3.TS.E2GAP,TS104039,TS104040,10.1145/3636534.3697325}.

\subsubsection{A1 Interface}connects and carries guidance from \ac{NonRTRIC} to \ac{NearRTRIC}~\cite{TS103983,TS103987,10024837}. In the data classification adopted in this paper, A1 is mainly associated with cross-domain data, especially \ac{EI} and policies, and with intelligence-derived data, especially rApp outputs~\cite{10298342,10225883,giannopoulos2026interoperable}.
A1 is realized as a \ac{REST}-based interface. In the current \ac{ORAN} specifications, \ac{A1AP} defines a \ac{REST} realization of A1 interface using \ac{HTTP} operations, while the underlying transport stack is based on \ac{TCP}/\ac{IP}, \ac{HTTP}/1.1 over \ac{TLS}, and \ac{JSON} as the data-interchange format. Accordingly, A1 mainly carries policy and enrichment-information objects represented in \ac{JSON}, with their relevant types defined in the A1 type-definition specification. Therefore, A1 is considered a guidance-oriented service interface rather than as a low-level control channel~\cite{TS103983,TS103987,TS103988}.

\subsubsection{O1 Interface} connects \ac{SMO} to \ac{NearRTRIC}, \ac{OCU}, \ac{ODU}, and, depending on the deployment model, \ac{ORU}. O1 is the primary interface through which management-plane data are collected from \ac{ORAN} functions towards \ac{SMO} and, indirectly, toward \ac{NonRTRIC} and its rApps~\cite{TS103982,O-RAN.WG10.TS.OAM,TS104043}. In particular, O1 is used to collect \ac{PM}, \ac{FM}, and \ac{CM} data (see Section ~\ref{sec:managmentdata}). In addition, O1 supports exposure of software-management data, file-management data, and records~\cite{TS104043}. Hence, O1 is the interface through which the \ac{SMO} configures \ac{ORAN} functions and collects management-level, historical, and supervisory data from them~\cite{11080542}.
Different O1 functions are based on different transfer mechanisms. O1 configuration procedures are based on \ac{NETCONF}, typically over \ac{SSH} or \ac{TLS}. O1 notifications are delivered as asynchronous \ac{JSON} messages over \ac{REST}/\ac{HTTPS}. For software and file management, \ac{NETCONF} \ac{RPC} procedures are used to initiate the operation, while the actual transfer may use \ac{FTPeS}, \ac{SFTP}, or \ac{HTTPS}~\cite{TS104043}.

\subsubsection{O2 Interface} connects \ac{SMO} to O-Cloud. It is the main interface through which cloud and deployment management data are exchanged between the orchestration framework and the infrastructure hosting virtualized \ac{ORAN} functions. In the \ac{ORAN} architecture, O2 provides two categories of services: O2 infrastructure management services (O2ims) and O2 deployment management services (O2dms)~\cite{O-RAN.WG6.TS.O2-GAP}. Accordingly, O2 is associated with O-Cloud management-plane data, including infrastructure inventory, infrastructure monitoring, alarm/fault data, configuration and provisioning data, software lifecycle information, and workload/deployment lifecycle information. In addition, O2 supports orchestration actions when \ac{SMO} updates, heals, or terminates \ac{ORAN} workloads in the cloud environment~\cite{TS103982,O-RAN.WG6.TS.O2-GAP}.

\subsubsection{E3 Interface} is an emerging interface proposed to support \ac{RT} interaction between dApps and \ac{RAN} nodes~\cite{LACAVA2025111342,10773685,boeira2026performance,crespo2025energy}. A recent dApp study~\cite{LACAVA2025111342} introduces E3 as an E2-like interface that enables dApps to obtain local \ac{RAN} data, perform \ac{RT} control actions. E3 interface supports procedures such as setup, subscription, indication, and control. A dApp first performs E3 setup procedure to authenticate and pair with the \ac{RAN} node. Then, it can subscribe to specific data or control capabilities exposed by the \ac{RAN} function. In addition, the same study proposes \textit{\ac{E2SMDAPP}}, a custom \ac{E2SM} for dApps, to expose dApp-generated inference results or reports to xApps through the E2 interface.

\subsubsection{R1 Interface} connects rApps to \ac{NonRTRIC} within \ac{SMO}. R1 is mainly associated with intelligence-derived data. It allows rApps to consume and produce R1 services, including service registration and discovery, authentication and authorization, data management and exposure, A1- and O2-related services, \ac{RAN} \ac{OAM}-related services, \ac{AIML} workflow services, and rApp management services~\cite{O-RAN.WG2.TS.R1GAP}.

\subsection{\ac{3GPP} Interfaces}
Besides \ac{ORAN} interfaces discussed above, the \ac{ORAN} architecture also uses interfaces defined by \ac{3GPP}. The \ac{ORAN} architecture description~\cite{TS103982,TS138401} includes E1, F1-C, F1-U, NG-C, NG-U, X2-C, X2-U, Xn-C, Xn-U, and Uu. However, from the data-centric perspective of this paper, the most relevant ones are E1, F1-C/F1-U, NG-C/NG-U, and Uu, because they carry the control-plane, user-plane, and core data.

\subsubsection{E1 Interface} connects control-plane functions at \ac{OCUCP} and user-plane functions at \ac{OCUUP}. E1 is used for exchange signalling and coordination data between them, and mainly associated with \ac{UE} context, bearer/session coordination context, and service/\ac{QoS} information~\cite{TS103982,TS138401,TS138460}.

\subsubsection{F1-C and F1-U Interfaces}
 F1-C interface is between \ac{OCUCP} and \ac{ODU}, while F1-U interface is between \ac{OCUUP} and \ac{ODU}. F1-C carries control-plane signalling between these entities and is realized by \ac{F1AP}, which supports procedures such as setup, reset, configuration update, and \ac{UE} context related signalling. In contrast, F1-U carries user-plane data between \ac{OCUUP} and DU. F1-C is mainly associated with control-plane and signalling data, whereas F1-U is mainly associated with user-plane traffic, bearer-level information, and service/\ac{QoS} data~\cite{TS103982,TS138401,TS38473}

\subsubsection{NG-C/N2 and NG-U/N3 interfaces}
NG-C interface, also referred to as N2, is between \ac{OCUCP} and \ac{AMF} in the core network, while the NG-U interface, also referred to as N3, is between \ac{OCUUP} and \ac{UPF}. NG-C/N2 carries control-plane data between the \ac{RAN} and the core, including \ac{UE}-context management, mobility-management signalling, paging, transport of \ac{NAS} messages, and \ac{PDU}-session-related procedures. NG-U/N3 carries user-plane traffic between \ac{OCUUP} and \ac{UPF}. Thus, NG-C/N2 is mainly associated with core network control and session context, whereas NG-U/N3 is mainly associated with user-plane tunnel context, packet/flow-level traffic observations, and service/\ac{QoS} data~\cite{TS103982,TS138401,TS38300}.

\subsubsection{Uu interface} is the radio interface between the \ac{UE} and \ac{ORAN} radio access functions. It is the point at which \ac{UE} traffic and signalling are carried over the air. Therefore, Uu is the origin of radio-domain data, \ac{PHY} measurements, \ac{UE} observations, and control plane signalling that is later processed by \ac{ORU}, \ac{ODU}, and \ac{OCU}~\cite{TS103982,TS38300}.

\subsubsection{Xn/X2 interfaces} \ac{ORAN} also relies on Xn and X2 interfaces for inter node coordination. Xn is between \ac{NGRAN} nodes, such as \acp{gNB}, while X2 is for LTE/E-UTRA nodes, such as \acp{eNB} or \ac{ORAN}-compliant O-eNBs~\cite{TS103982,TS138401,TS138420}. These interfaces support mobility, dual connectivity, and inter-node control/user-plane data exchange, and their messages may be exposed to the \ac{NearRTRIC} through \acp{E2SM} such as \ac{E2SMNI}~\cite{ORAN-WG3.E2SM-NI}.

\subsection{Agentic \ac{AI} Interoperability Interfaces} 

\label{subsec:agentic_interfaces}

Besides \ac{ORAN}/\ac{3GPP} interfaces, Agentic \ac{AI} applications may require additional \acp{API} for tool discovery, context sharing, and reasoning coordination.

\subsubsection{\ac{MCP}}is a \ac{JSON}-\ac{RPC} client-server protocol for connecting \ac{AI} applications to external data sources and tools~\cite{MCP_Intro}. It is often described as the \textit{"USB-C for \ac{AI}"}~\cite{ehtesham2025survey}. In \ac{ORAN}, \ac{MCP} serves two roles: 
(i) \textit{Data Acquisition (Resources):} \ac{MCP} standardizes how applications expose contextual data to \acp{LLM} as \emph{Resources}, which in an \ac{ORAN} setting may include live \ac{KPM} streams, configuration files, logs, or other operational datasets used to ground the model's reasoning; (ii) \textit{Action Delivery (Tools):} \ac{MCP} standardizes how applications expose executable capabilities as \emph{Tools}, enabling the \ac{LLM} to invoke specific operations through structured tool calls, for example querying telemetry, validating configurations, or triggering selected network-management functions.

\subsubsection{\ac*{ACP}} \ac{ACP}is a \ac{REST}-native protocol for structured communication among heterogeneous agents~\cite{ibm_acp_2025}. It supports multipart messages and both synchronous and asynchronous interactions, making it suitable for scalable \ac{ORAN} orchestration where agents built on different frameworks must exchange session-aware metadata and instructions.

\subsubsection{\acs*{A2A}} \ac{A2A} supports peer collaboration through concepts such as agent cards, tasks, messages, and artifacts~\cite{google_a2a_2025}. In \ac{ORAN}, this is used for coordinating complex tasks across multiple domains, where a "Client Agent" (such as an intent manager) can discover and invoke the specialized "Skills" of "Remote Agents" (such as an energy-saving rApp or a traffic-steering xApp).

\subsubsection{\ac*{ANP}} \ac{ANP} is a decentralized, peer-to-peer framework designed for collaboration over the open internet using \ac{W3C} \acp{DID}~\cite{anp_github}. \ac{ANP} provides the foundation for future federated 6G ecosystems, allowing agents to autonomously discover and authenticate peers across different administrative or vendor domains without centralized intermediaries

\begin{table*}[t]
\centering
\renewcommand{\arraystretch}{1.6}
\caption{Statistical Applications in \ac{ORAN}}
\label{tab:Statistical-Applicationa}

\begin{tabular}{|p{0.9cm}|p{1.5cm}|p{4.4cm}|p{3.8cm}|p{1cm}|p{3.9cm}|}
\hline
\rowcolor{lightgray}
\multicolumn{1}{|c|}{\textbf{Ref.}} &
\multicolumn{1}{c|}{\textbf{Deployment}} &
\multicolumn{1}{c|}{\textbf{Objective}} &
\multicolumn{1}{c|}{\textbf{Data}} &
\multicolumn{1}{c|}{\textbf{Interface}} &
\multicolumn{1}{c|}{\textbf{Method}} \\
\hline

~\cite{10617647} &
rApp &
Detection of configuration conflicts&
\acp{KPM} and \ac{CM} data &
O1 &
Clustering and Pearson correlation \\
\hline

~\cite{10.1145/3721889.3721927} &
rApp &
Energy-aware edge resource orchestration &
Workload, latency, computing-resource state &
O1 &
Heuristic optimization \\
\hline

~\cite{10226043} &
xApp &
Detection of signaling storm attacks &
\ac{RRC} Setup Request messages, \acp{KPM} &
E2 &
Statistical profiling, thresholding \\
\hline

~\cite{9877931} &
xApp &
\ac{QoE}-driven user association, resource allocation, power allocation &
\acp{KPM} &
E2 &
Adaptive Genetic Algorithm \\
\hline

~\cite{crespo2025energy} &
dApp &
Energy-aware \ac{CPU} management &
OS/host telemetry &
E3 &
Rule-based heuristic control \\
\hline

~\cite{10773685} &
dApp &
\ac{RT} spectrum sharing &
\ac{IQ} samples &
E3 &
Energy-threshold sensing \\
\hline

~\cite{11044706} &
Hierarchical &
Channel-aware energy saving &
Radio measurements, RU state, O-Cloud workload data &
A1, E2 &
\ac{BILP}-based optimizer \\
\hline

\end{tabular}
\end{table*}

\section{Abstraction View of \ac{ORAN} Data}
\label{sec:Abstraction}
The abstraction view describes how \ac{ORAN} data are represented to be consumed by applications. In this section, we discuss each abstraction level in terms of data types, production tools, applications, and abstraction limitations.

\subsection{Raw Data}
Raw data refer to measurements, or observations collected with little processing. This includes \ac{IQ} samples, \ac{RF} measurements, pcap and log files. Such data preserve high fidelity and can be captured using \ac{SDR} tools, packet analyzers, logging systems, and monitoring mechanisms~\cite{10.1145/3570361.3592493}. Raw data are useful for applications requiring fine-grained visibilit. However, it is expensive to store, and transport, and may expose privacy concerns~\cite{8428412}.

\subsection{Aggregated and Statistical Data}
Aggregated data refer to measurements summarized over time windows, users, cells, slices, or flows. Statistical data capture the distribution of these measurements (such as \acp{KPM}). These data can be produced using E2 service-model reporting and statistical analysis tools. Aggregated and statistical data are suitable for applications that require a compact view of network state~\cite{9814869}. However, aggregation may hide short term events, and anomalies. Wen \emph{et al.}~\cite{10.1145/3565474.3569070} make this point by arguing that \ac{ORAN} \ac{KPM} is insufficient for robust security analysis and by proposing other fine-grained telemetry.

\subsection{Derived and Enriched Data}
Derived data refer to features computed from one or more measurements, while enriched data add external context. These data, such as congestion scores, traffic asymmetry, can be produced using feature engineering pipelines, and are useful when applications require more than isolated telemetry values. LS \emph{et al.}~\cite{11152804} illustrate this by jointly using \ac{RAN} telemetry, platform telemetry, and infrastructure telemetry to predict faults in virtualized \ac{ORAN} deployments. Management Data Analytics also supports this abstraction level by transforming management data into analytics that can enhance \ac{SMO} and \ac{NonRTRIC} capabilities~\cite{10604823}. The main limitation is that the quality of this abstraction depends on feature selection.

\subsection{Intelligence-Derived Data} Intelligence-derived data refer to outputs produced by all types of \ac{ORAN} applications(rule-based, \ac{MLDL}, and \ac{AI} agentic). This includes predictions, classifications, recommended actions, and feedback from previous decisions. Since these outputs may later be reused by other applications, they should be treated as data objects rather than final decisions only. Therefore, intelligence-derived data may face limitations related to validation, timestamping and confidence assessment before being used in other control tasks~\cite{kim2026development}.

\section{Application View of \ac{ORAN} Data}
\label{sec:Applications}

\ac{ORAN} applications consume data differently depending on their objective, location, and control timescale. In this section, we classify applications according to how they process data and generate decisions: statistical applications, \ac{MLDL}-based applications, and Agentic \ac{AI} applications. This view shows how \ac{ORAN} data are used across different levels of intelligence.

\subsection{Statistical Applications in \ac{ORAN}}
\label{sec:TraditionalApps}

Not all intelligent \ac{ORAN} applications rely on \ac{AI}, a part of applications uses instead traditional statistical, heuristic, or optimization methods, such as, clustering, correlation analysis, and others. These methods often require less data, and can operate with lower computational cost. We organize these applications according to their location of deployment (Table\ref{tab:Statistical-Applicationa}).
\subsubsection{rApp} 
Armstrong \emph{et al.}~\cite{10617647} introduce a conflict detection rApp. The system leveraged \ac{KPM} and configuration data, using clustering and Pearson correlation to identify cells with conflictive configuration conditions.
Levis \emph{et al.}~\cite{10.1145/3721889.3721927} present SLEEPY-rApp, an intelligent orchestration rApp for resource allocation in \ac{ORAN}. SLEEPY-rApp models a joint optimization problem over server activation status, request routing, and computing allocation across edge servers co-located with DUs/RUs. Since the resulting problem is NP-hard, the authors propose a low-complexity heuristic that periodically reconfigures the system according to workload and latency conditions.

\subsubsection{xApp} 
The paper~\cite{10226043} proposes xApp to detect signaling storm attacks. It uses copies of Random Access Response messages containing \ac{TA} parameter values and \ac{RRC} Setup Requests messages, and historical \ac{KPM} profiles of the number of RSRs associated with a given \ac{TA} value. The xApp identifies \ac{IoT} devices by using \ac{TA}, then monitors the number of RSRs associated with each \ac{TA} and compares it with \ac{KPM} profiles to detect anomalies. Another xApp is proposed in~\cite{9877931} to optimize \ac{QoE}-driven user association, resource allocation, and power allocation. The solution consumes \acp{KPM}, including transport block size index, base station capacity indicators, \ac{UE} service demand information, and throughput metrics. The optimization problem is formulated as a dual 0/1 Multi-Knapsack Problem and solved using an Adaptive Genetic Algorithm combining genetic evolution with simulated annealing.

\subsubsection{dApp} The paper~\cite{crespo2025energy} proposes a lightweight dApp for \ac{CPU} management within the \ac{ODU}. The dApp relies on rule-based heuristics and optimization principles to dynamically adjust \ac{CPU} affinity, thread isolation, and frequency scaling based on host telemetry. 
Gangula \emph{et al.}~\cite{10773685} propose a dApp for \ac{RT} spectrum-sharing that uses an energy-threshold sensing method on locally extracted \ac{IQ} samples to detect incumbent-user activity and trigger \ac{PRB} barring at base station.

\subsubsection{Hierarchical}~\cite{11044706} proposes GreenRAN, an \ac{ORAN} framework integrating an rApp and xApp for energy-aware resource management. The rApp, \ac{CARC} dynamically deactivates underutilized radio units based on \ac{RT} channel and traffic data, while the xApp, \ac{CDPM}, consolidates workloads across edge nodes using a greedy \ac{BILP}-based optimizer.

\subsection{\ac{MLDL} Applications in \ac{ORAN}} \label{Data Analysis methods}
Recent \ac{ORAN} research leverages \ac{MLDL} models in network management. \ac{MLDL}-based applications exploit heterogeneous data sources including \ac{KPM}, fine-grained \ac{RAN} measurements, \ac{IQ} samples, O2 cloud statistics, and even traffic flow data. Unlike statistical methods that rely on predefined thresholds or correlation rules, \ac{MLDL} models can capture nonlinear relationships of network data.
We organize \ac{MLDL} applications according to their deployment location (Table~\ref{tab:MLDL-Applications}). 

\subsubsection{rApp} 
In~\cite{10620743}, rApps are deployed to support interference reduction. The framework includes four rApps for localization, \ac{UE} throughput prediction, predictive maintenance, and interference management. These rApps use \acp{KPM} such as \ac{RSRP}, \ac{RSRQ}, \ac{SINR}), \ac{RSSI}, \ac{MCS}, \ac{PRB} usage, and throughput. Tran \emph{et al.}in ~\cite{10620743} evaluate both classical \ac{ML} and \ac{DL} models, including \ac{SVM}, Random Forest, \ac{XGBoost}, \ac{DNN}, and \ac{LSTM}, and show that sequence-based models achieve the best results. In~\cite{kim2026development},Kim \emph{et al.} present a framework for the development and functional validation of \ac{MLDL} rApps in \ac{ORAN}. The framework supports the full rApp lifecycle, from data generation and model training to integration, onboarding, and closed-loop validation. To demonstrate its utility, the authors implement a network energy-saving rApp, using an \ac{LSTM} model to predict future \ac{PRB} utilization and \ac{UE} counts and trigger cell on/off decisions. On the other hand, Vilà \emph{et al.}~\cite{10008506} implement a capacity sharing rApp for \ac{ORAN} network slicing. The \ac{DQNMARL}-based rApp gathers \ac{PM} data from the \ac{ODU}, and configures slice resource allocation ratio through \ac{NETCONF}~\cite{5560601}. The rApp dynamically adjusts slice capacity to match traffic demands while preserving \acp{SLA}.

\subsubsection{xApp}~\cite{10597083} develops a \ac{LSTM}- based method in \ac{FL} settings for anomaly detection (\ac{DDoS} attacks) and identification of network slices. The xApps used \acp{KPM} data of a network digital twin of downlink/\ac{UL} of E2 nodes: samples transmitted; queue length; throughput; number of packets transmitted. On the other hand, Groen \emph{et al.} in~\cite{10622798} use 17 \ac{ORAN} \acp{KPM} to classify \ac{UE} traffic into four categories: \ac{eMBB}, \ac{URLLC}, and \ac{mMTC} and control data using \ac{CNN}-based xApp. The paper~\cite{11162211} introduces an xApp (\ac{CNN}-based) integrated within \ac{NearRTRIC} for dynamic network slicing and resource optimization in \ac{ORAN}. Using data from a simulated ns-O-RAN environment: throughput, delay, \ac{PER}, \ac{GBR}, \ac{SINR}), they trained multiple \ac{ML} models, including \ac{ANN} and \ac{SVM}, to classify users into \ac{eMBB}, \ac{URLLC}, and \ac{mMTC} slices.
Moreover, the paper~\cite{10849726} uses \ac{KPM} to detect signaling storm attacks,It evaluates several \ac{MLDL}, including \ac{BiDLSTM}, \ac{LSTM}, \ac{SVM}, Random Forest, \ac{DT}, and \ac{KNN}. Also, He\emph{et al.} ~\cite{10978529} use \ac{KPM} and \ac{MLP} to predict slice of traffic. 
 
The paper~\cite{xavier2024crossdomainaiearlyattack} went further and used beside \acp{KPM} data and the network traffic flow to train \ac{AI} model for early detecting of malicious flows in \ac{ORAN}. They introduced the idea of using cross-domain \ac{AI} to train the classifier model using \ac{KPM} datasets mapped to flow datasets.

Using \ac{IQ} samples, the paper~\cite{Systemlevel} targeted the detection of poisoning attacks by training \ac{CNN}-based xapp to classify whether the network is experiencing interference generated by the jammer attacker or not. They also considered using an adaptive \ac{MCS} approach to mitigate these types of attack (switching between adaptive or fixed \ac{MCS}).

Several works exploit \ac{PHY} metrics to enable intelligent security in \ac{ORAN}. These approaches rely on measurements such as \ac{RSRP}, which reflect radio conditions and can reveal inconsistencies caused by malicious transmitters or rogue base stations. For instance, Huang \emph{et al.}~\cite{10225868} developed an xApp for rogue base station detection in \ac{ORAN} testbed. The approach leverages \ac{UE} reported \ac{PHY} measurements (signal strength data from the base station to differentiate the rough base station from the operational base station), including: \ac{RSRP}, \ac{SSB} signal strength, and \ac{PCI}, from which signal stability features (standard deviation over time) are derived. Supervised \ac{ML} models (Random Forest, \ac{KNN}, and \ac{SVM}) are trained at \ac{NearRTRIC} using aggregated \ac{PHY} measurements, and then are redistributed through the \ac{RAN} to enable distributed detection at the \ac{UE} side as Android application. On the other hand, Orhan \emph{et al.} in~\cite{9679991} propose an xApp for intelligent connection management that leverages \ac{PHY} measurements to optimize user–cell association. The primary input data consist of \ac{RSRP} measurements collected from serving and neighboring cells. From these measurements, the xApp derives additional radio features including estimated channel capacity (based on \ac{RSRP} and a noise model), \ac{UE} achievable data rate, and cell load (number of connected \acp{UE}). The \ac{RAN} is abstracted as a graph where \acp{UE} and cells are modeled as nodes and wireless links are weighted using \ac{PHY}-derived metrics. A \ac{GNN} combined with Deep Q-Learning is trained and deployed to learn optimal association decisions. The objective (based on the reward function) can be configured to maximize sum throughput, improve cell performance, or achieve load balancing. 

The paper~\cite{10283501} proposed a semi-supervised \ac{DAE} xApp for \ac{RF} anomaly detection in \ac{ORAN}. Using a 10,000 sample dataset with features such as \ac{RSRP}, \ac{RSRQ}, \ac{PRB} utilization, and throughput, the \ac{DAE} was trained to reconstruct normal \ac{RF} patterns and detect anomalies based on reconstruction loss. On the other hand, the paper~\cite{11080593} demonstrated a \ac{QoSBRO} xApp deployed on a multi-vendor \ac{ORAN} testbed. Using historical and live telemetry data, including \ac{PRB} usage, \ac{SINR}), \ac{RSRQ}, and \ac{QoS} metrics, from VIAVI’s RAN scenario generator, the xApp employed \ac{LSTM} and Random Forest models to predict and adjust \ac{PRB} allocations across network slices.

\subsubsection{dApp} In~\cite{olimpieri2025libiq}, Olimpieri \emph{et al.} propose LibIQ, a dApp for \ac{RT} \ac{RF} spectrum classification in \ac{ORAN}. The proposed dApp operates directly alongside the DU/CU and processes \ac{IQ} samples locally. LibIQ provides analysis and preprocessing functions, including spectrogram generation, and an Energy Peak Detector, and embeds a \ac{CNN} classifier inside the dApp for multiclass signal identification. The work in~\cite{verbruggen2026disaggregated} proposes OSIRIS, a lightweight \ac{DL} model for interference classification in \ac{ORAN}. The method exploits Split 7.2x disaggregation by extracting features at \ac{ORU} and frequency/CSI features at \ac{ODU}, which are then fused for final classification. The solution is explicitly motivated by \ac{RT} dApp deployment, where interference sensing and mitigation must meet strict latency constraints. Lacava \emph{et al.}~\cite{LACAVA2025111342} present a reference architecture for dApps in \ac{ORAN}, enabling \ac{RT} \ac{AI}-based control directly at the \ac{OCU}/\ac{ODU}. The paper introduces E3 interface, defines dApp lifecycle management and coordination with xApps through \ac{E2SM} for dApps (E2SM-DAPP), and validates the framework through \ac{OAI}-based experiments showing sub millisecond control-loop latency for spectrum sharing and positioning use cases. Moreover, Neasamoni Santhi \emph{et al.}~\cite{10.1145/3704413.3764454} present dApp for \ac{RT} \ac{UL} interference detection in \ac{ORAN}. Specifically, Neasamoni Santhi \emph{et al.} in~\cite{10.1145/3704413.3764454} design a \ac{GPU}-accelerated dApp that runs in \ac{gNB} physical layer and uses \ac{CNN} model to process raw \ac{IQ} samples together with \ac{PHY} metrics such as \ac{RSSI}, \ac{RSRP}, \ac{SINR}), \ac{MCS} and code-block error statistics. 

\begin{table*}[!t]
\centering
\renewcommand{\arraystretch}{1.6}
\caption{\ac{MLDL} Applications in \ac{ORAN}}
\label{tab:MLDL-Applications}

\begin{tabular}{|p{0.8cm}|p{1.5cm}|p{4cm}|p{4.3cm}|p{1.9cm}|p{3.8cm}|}
\hline
\rowcolor{lightgray}
\multicolumn{1}{|c|}{\textbf{Ref.}} &
\multicolumn{1}{c|}{\textbf{Deployment}} &
\multicolumn{1}{c|}{\textbf{Objective}} &
\multicolumn{1}{c|}{\textbf{Data}} &
\multicolumn{1}{c|}{\textbf{Interface}} &
\multicolumn{1}{c|}{\textbf{\ac{MLDL}}} \\
\hline

~\cite{10620743} &
rApp &
Interference reduction in real-world 5G deployment &
Management-plane \ac{PM}/\ac{KPM} data and radio measurements &
O1 / R1 &
\ac{SVM}, Random Forest, \ac{XGBoost}, \ac{DNN}, \ac{LSTM} \\
\hline

~\cite{kim2026development} &
rApp &
Network energy saving &
\ac{PM} data: \ac{PRB} utilization and \ac{UE} count &
O1 / R1 / A1 &
\ac{LSTM} \\
\hline

~\cite{10008506} &
rApp &
Capacity sharing for \ac{RAN} slicing &
Slice-level \ac{PM}/\ac{KPM} data from \ac{ODU} &
O1 &
\ac{DQNMARL} \\
\hline

~\cite{10597083} &
xApp &
Anomaly detection and slice identification &
\ac{KPM} data from \ac{ORAN} digital twin &
E2 &
Federated \ac{LSTM} \\
\hline

~\cite{10622798} &
xApp &
Traffic slice classification &
\acp{KPM} &
E2 &
\ac{CNN} \\
\hline

~\cite{11162211} &
xApp &
Network slicing and resource optimization &
Slice-level \ac{QoS}/\acp{KPM}&
E2 &
\ac{SVM}, Random Forest, \ac{ANN}, \ac{CNN} \\
\hline

~\cite{10849726} &
xApp &
Signaling storm attack detection &
\acp{KPM} &
E2 &
\ac{BiDLSTM}, \ac{LSTM}, \ac{SVM}, Random Forest, \ac{DT}, \ac{KNN} \\
\hline

~\cite{10978529} &
xApp &
Distortion-tolerant slice prediction &
\acp{KPM} &
E2 &
\ac{MLP} \\
\hline

~\cite{xavier2024crossdomainaiearlyattack} &
xApp &
Early malicious-flow detection &
\acp{KPM} and network traffic flows &
E2 &
Cross-domain \ac{ML} classifier \\
\hline

~\cite{Systemlevel} &
xApp &
Adversarial attack analysis and defense for xApps &
spectrograms and \acp{KPM} &
xApp &
\ac{CNN}, \ac{DNN}, knowledge distillation \\
\hline

~\cite{10225868} &
xApp &
Rogue base-station detection &
Fine-grained \ac{PHY} measurements: \ac{RSRP}, \ac{SSB} strength, \ac{PCI} &
xApp/\ac{UE} app &
Random Forest, \ac{KNN}, \ac{SVM} \\
\hline

~\cite{9679991} &
xApp &
User--cell association and load balancing &
Fine-grained radio measurements and graph features &
E2 &
\ac{GNN}, Deep Q-Learning \\
\hline

~\cite{10283501} &
xApp &
\ac{RF} anomaly detection &
Fine-grained \ac{RAN} telemetry: \ac{RSRP}, \ac{RSRQ}, \ac{PRB}, throughput &
\ac{NearRTRIC} / xApp &
\ac{DAE} \\
\hline

~\cite{11080593} &
xApp &
\ac{QoS}-based \ac{PRB} optimization &
Slice-level \ac{QoS} data and \ac{PRB}-allocation data &
E2&
\ac{LSTM}, Random Forest \\
\hline

~\cite{olimpieri2025libiq} &
dApp &
\ac{RT} \ac{RF} spectrum classification &
Radio-level data: \ac{IQ} samples and spectrograms &
dApp / local \ac{RAN} access &
\ac{CNN} \\
\hline

~\cite{verbruggen2026disaggregated} &
dApp &
Cross-technology interference classification &
\ac{ORU} time samples, \ac{ODU} frequency symbols, CSI &
\ac{OFH}/local access &
Lightweight \ac{DL} with multi-domain fusion \\
\hline

~\cite{LACAVA2025111342} &
dApp &
\ac{NearRT} dApp-based \ac{RAN} control &
Radio-level and user-plane data: \ac{IQ} samples, \ac{CIR}, \ac{PHY} data &
E3&
\ac{RT} dApp inference framework \\
\hline

~\cite{10.1145/3704413.3764454} &
dApp &
\ac{RT} \ac{UL} interference detection &
Radio-level data: \ac{UL} \ac{IQ} samples &
E3 &
\ac{GPU}-accelerated \ac{CNN} \\
\hline

~\cite{10078092} &
Hierarchical &
Communication and computation slicing for \ac{URLLC} &
\ac{RAN} state and \ac{MEC}-computation state &
O1/E2 &
\ac{RL} \\
\hline

~\cite{9927252} &
Hierarchical &
Predictive closed-loop resource provisioning for slicing &
Slice-level \ac{PM}/\ac{KPM} data and O-Cloud statistics &
O1 / O2 / A1 / E2 &
\ac{LSTM} \\
\hline

~\cite{10757839} &
Hierarchical &
Energy saving and traffic steering coordination &
\ac{PM} data and traffic-load data &
A1, E2 &
\ac{ML}-based traffic-load prediction \\
\hline

~\cite{9367582} &
Hierarchical &
Handover optimization &
Management-plane \ac{PM}/\ac{CM} data: \ac{NCRT} and handover statistics &
O1 / A1 &
Random Forest \\
\hline

~\cite{11260483} &
Hierarchical &
\ac{ORAN} resource management and slicing &
Slice state, RB allocation, DU-level resource state &
RIC / E2 &
Multi-Agent \ac{RL} \\
\hline

\cite{ddos} & Hierarchical & Multi-layer \ac{DDoS} detection and mitigation & 
\ac{UE} traffic, \ac{RAN} telemetry, cross-\ac{RAN} information, service usage data & 
E2 & 
Random Forest, \ac{SVM}, \ac{LR}, \ac{KNN}, \ac{DT}, Naive Bayes, \ac{AdaBoost}, \ac{QDA}, Extra Trees, Isolation Forest, and \ac{FNN} \\
\hline

\cite{PQC} & Hierarchical & Energy-aware \ac{PQC} scheduling & Cell load, latency headroom, mobility statistics, cryptographic policy, energy state & A1 / O1 / E2 & Constrained \ac{RL} \\
\hline

\end{tabular}
\end{table*}

\subsubsection{Hierarchical} Filali \emph{et al.} in~\cite{10078092} propose a deep \ac{RL}–based xApp for joint communication and computation slicing targeting \ac{URLLC} services. The solution performs offline training in the \ac{NonRTRIC} using O1-collected \ac{PM} data, while the \ac{NearRTRIC} collects online environmental state information (e.g., channel gain, \ac{PRB} utilization, computing availability) via E2 interface. The trained deep \ac{RL} agent produces slicing control decisions (radio block and \ac{CPU} allocation) delivered via E2, forming a hierarchical closed-loop control architecture. On the other hand, the paper~\cite{9927252} introduces a predictive resource provisioning framework for \ac{ORAN} network slicing, where \ac{AI} is integrated into both the \ac{NonRTRIC} and \ac{NearRTRIC} to proactively prevent \acp{SLA} violations. The system collects slice-level \acp{KPM} such as number of active users per bearer, traffic volume, and downlink \ac{PRB} utilization via the O1 interface, as well as cloud resource statistics via O2. These measurements are used to train an \ac{LSTM}-based recurrent neural network that predicts future slice resource demand. The trained model runs as an rApp in the \ac{NonRTRIC}, generating slice descriptors and policies that are sent via A1 to the \ac{NearRTRIC}. The \ac{NearRTRIC} (via xApps) dynamically scales radio resources (\acp{PRB}), while cloud resources are scaled through O2. This work is one of the early demonstrations of \ac{AI}-driven predictive closed-loop slicing automation in \ac{ORAN}, combining both \ac{RAN} and cloud resource orchestration. In addition, the paper~\cite{10757839} demonstrated a multi-vendor interoperable \ac{ORAN} integrating an \ac{AI}-based \ac{ES} rApp with a Traffic Steering xApp to resolve operational conflicts between energy efficiency and service quality. Validated using a commercial Vodafone dataset, the \ac{ES} rApp utilizes \ac{ML} algorithms to predict traffic load and determine optimal ON/OFF transitions for capacity carriers. To maintain user experience during these transitions, the framework employs two coordination mechanisms: A1 interface policies with "FORBID" clauses to steer traffic away from candidate cells, and \ac{E2SMCCC} service model, which allows Traffic Steering xApp to autonomously offload users upon detecting an energy saving state transition. An automatic neighbor relation optimization framework for \ac{ORAN} was proposed in~\cite{9367582}. The solution leverages \ac{PM} data collected from \ac{RAN} nodes via O1 interface and stored in an \ac{SMO}-managed dataset, which includes attributes such as Network Element identifiers, event timestamps, handover failure causes, and Neighbor Cell Relation Table statistics. A supervised Random Forest classifier is trained in \ac{NonRTRIC} to predict potential handover failures and proactively determine whether specific neighbor cells should be added, removed, or temporarily restricted. The trained model is deployed as an rApp in \ac{NonRTRIC}, and its decisions are translated into A1 policies sent to a \ac{NearRTRIC} xApp for \ac{RT} enforcement. This architecture implements a hierarchical two-level control loop combining long-timescale predictive optimization and \ac{NearRT} execution. The work in \cite{11260483} proposes a hierarchical \ac{ORAN} resource management framework for joint slicing and scheduling, where DU-local agents (dApps) learn resource-allocation policies and a \ac{NearRTRIC} xApp acts as a global critic. Soleymani \emph{et al.}~ \cite{ddos} designed a hierarchical \ac{ML}-based security framework for \ac{ORAN} to detect and mitigate \ac{DDoS} attacks. The framework distributes intelligence such as \ac{LR}, Naive Bayes, \ac{AdaBoost}, \ac{QDA}, Extra Trees, Isolation Forest, and \ac{FNN} models, across multiple layers by deploying a lightweight dApp in the \ac{RAN} for sub-10 ms anomaly detection, an xApp-U in the \ac{NearRTRIC} for more comprehensive cross-\ac{RAN} anomaly analysis, and an xApp-S for tracking the service usage of suspicious \acp{UE}.
Moreover, Gupta \emph{et al.}~\cite{PQC} designed an energy-aware framework for integrating \ac{PQC} into \ac{ORAN} while reducing the energy and latency overhead associated with \ac{PQC} handshakes. The framework adopts a hierarchical \ac{ORAN} architecture in which a Crypto Policy rApp in the \ac{NonRTRIC} defines long-term cryptographic policies, while a \ac{SOS} xApp in the \ac{NearRTRIC} employs constrained \ac{RL} to optimize the scheduling and placement of cryptographic operations.

\subsection{Agentic \ac{AI} Applications in \ac{ORAN}} 

Agentic \ac{AI} in \ac{ORAN} goes beyond using an \ac{AI} model for one inference~\cite{aghayev2026genesis}. Following~\cite{sapkota_ai_2026}, it refers to a coordinated set of \ac{AI} agents that can interpret intents, reason over context, use tools, and adapt actions. These agents may operate individually for a specific \ac{RAN} task or coordinate with other agents across different \ac{ORAN} control loops~\cite{navidan2026toward,elkael2025agentran}
Agentic \ac{ORAN} applications rely on various types of data such natural-language intents, telemetry, logs and others to support coordinated closed-loop operations.

We categorize Agentic \ac{AI} \ac{ORAN} applications into three groups: (i) single-agent control, in which one \ac{LLM} agent supports one \ac{RAN} task; (ii) multi-agent agentic systems operating mainly within one control domain (\ac{NonRT}/\ac{NearRT} level); and (iii) hierarchical multi-timescale multi-agent agentic frameworks spanning the \ac{NonRT}, \ac{NearRT}, and \ac{RT} levels. Table~\ref{tab:agentic_oran_llm_summary} summarizes \acp{LLM} used in these Agentic \ac{AI} applications and their usage strategies.

\subsubsection{ Single-Agent Applications in one \ac{ORAN} control Loop}
This group illustrates the transition from \ac{LLM}-assisted control toward using richer forms of Agentic \ac{AI} in \ac{ORAN}, where \acp{LLM} begin to move from decision support to autonomous control. For example, LLM-xApp~\cite{osti_10628444} is an early example of intent-driven radio resource management in \ac{ORAN}. The xApp uses slice telemetry, and applies an \ac{OPRO}~\cite{ yang2023large}, in which \ac{GPT}-4 iteratively generates \ac{PRB} allocation decisions from structured meta-prompts. To avoid the need for offline training or fine-tuning, the agent uses In-Context Learning and refines its strategy using an optimization trajectory (a sorted history of past performance scores) included in its meta-prompts. Furthermore, a decaying temperature mechanism is employed to balance the initial exploration of diverse allocation strategies with the eventual exploitation of high-utility solutions. Beyond \ac{LLM}- based control, Feng \emph{et al.} in~\cite{feng2025towards} present a high-level semantic–Agentic \ac{AIRAN} framework for 6G, where semantic/task data and radio/control data are jointly used to enable \textit{autonomous control}. The core method is a two-timescale design that combines semantic communication with \ac{RL}/\ac{MARL}-based agentic control: a fast loop performs online \ac{RAN} decision-making from current radio/control observations, while a slower loop updates semantic models, policies, and high-level knowledge used to guide future decisions. The paper also discusses \ac{LLM}-based agents and knowledge-graph reasoning as high-level planning and coordination mechanisms. The work in~\cite{bao2025llm} introduces \ac{LLM}-hRIC, where an \ac{LLM}-based rApp in the \ac{NonRTRIC} generates strategic guidance from global multimodal network data, including long-term performance metrics, alarm logs, topology data, radio-map information, and average channel/interference statistics, while a \ac{RL}-based xApp in \ac{NearRTRIC} refines this strategic guidance using local \ac{RT} observations such as channel gain, user load, and local rate data.

\subsubsection{Multi-Agent Agentic \ac{AI} Applications in one \ac{ORAN} control Loop}
This group moves beyond single-agent control and introduces multiple agents that cooperate within one control loop, usually at the \ac{NonRTRIC} or \ac{NearRTRIC} level. For example, Bimo \emph{et al.}~\cite{bimo2025intent} present an intent-based \ac{RAN} management framework that uses \acp{LLM} to translate formalized high-level intents into executable \ac{ORAN} configuration strategies through a closed-loop process. Their system combines a strategist agent for intent translation and strategy generation with a history analyzer agent that uses past attempts and real time \ac{PM} observations to refine subsequent decisions, while relying on the O1 interface for \ac{PM} feedback and configuration updates. Moreover, Bimo \emph{et al.} in~\cite{bimo2026contract} propose \ac{CAIF} for \ac{ORAN} network slicing, in which natural-language operator intents are interpreted and validated by a dual-agent \ac{LLM} pipeline (Profiling Agent and Evaluator Agent) before being transformed into a formal Intent Contract via the TMF-921 Intent Management \ac{API}~\cite{TMF921}. Rather than allowing the \ac{LLM} to directly produce executable \ac{RAN} actions, the framework delegates execution to a hierarchical rApp/xApp workflow: the \ac{SLA} Slice rApp and A1 Policy Handler rApp process the validated contract at the \ac{SMO}/\ac{NonRTRIC}, while the KPIMON-GO xApp, \ac{SLA} Slice xApp, and RC xApp realize the closed-loop control process at the \ac{NearRTRIC}.

Chatzistefanidis \emph{et al.}~\cite{chatzistefanidis2025mx} proposed MX-AI, an agentic observability and control platform for Open and \ac{AIRAN}, where a multi-agents \ac{LLM} is deployed at \ac{SMO} on R1 interface to process natural-language operator intents. The framework first understands the operator’s request, then gathers fresh network context from live O1/E2 telemetry, logs, topology information, and \ac{CRD} state, and finally either provides observability answers or enforces the required control action(such as FlexRIC policy objects, or A1 policy updates), which are then applied through deployment tools over the O1/A1 interfaces to support closed-loop \ac{RAN} reconfiguration. In addition, the paper~\cite{sharma2025mobillm} presents MobiLLM, an Agentic \ac{AI} framework for threat mitigation in \ac{ORAN} that can be instantiated as an rApp within the \ac{SMO}/\ac{NonRTRIC} and operate with existing threat-detection xApps and rApps. It consists of three main agents: a Threat Analysis Agent that triggers alerts using related network data, such as \ac{PM} metrics, RAN-level data, and logs; a Threat Classification Agent that uses \ac{RAG} to map the analyzed incident to the MITRE FiGHT threat base~\cite{MITRE_FiGHT} and extract the corresponding mitigation guidance, while grounding its reasoning in trusted sources such as \ac{3GPP} specifications, \ac{ORAN} documents, and curated academic literature; and a Response Planning and Execution Agent that translates the selected mitigation into safe \ac{API}-based actions over \ac{ORAN} control interfaces, subject to human approval.

\begin{table*}[h]
\centering
\renewcommand{\arraystretch}{1.6}
\caption{Summary of \ac{LLM} Usage in Agentic \ac{AI} \ac{ORAN} Applications}
\label{tab:agentic_oran_llm_summary}

\begin{tabular}{|p{0.8cm}|p{7cm}|p{8cm}|}
\hline
\rowcolor{lightgray}
\multicolumn{1}{|c|}{\textbf{Ref.}} &
\multicolumn{1}{c|}{\textbf{Used Models}} &
\multicolumn{1}{c|}{\textbf{Usage Strategy}} \\
\hline

\cite{bimo2026contract} &
Profiling Agent: Qwen3-4B-Instruct-2507~\cite{qwen2025qwen3, qwen2025qwen3_4b_instruct_2507}; Evaluator Agent: Llama-3.3-Nemotron-Super-49B-v1.5~\cite{nvidia2025llama33nemotronsuper49b15} &
Prompt-based iterative intent translation (no fine-tuning) \\
\hline

\cite{chatzistefanidis2025mx} &
\ac{GPT}-4.1~\cite{openai2025gpt41}; \ac{GPT}-4.1-mini~\cite{openai2025gpt41mini}; llama3.3~\cite{meta2024llama33_70b}; qwen2~\cite{chu2024qwen2} &
\ac{RAG} with tool grounding and live push/pull context refresh (no fine-tuning). \\
\hline

~\cite{elkael2025agentran} &
Claude Sonnet 4~\cite{anthropic2025introducingclaude4} &
In-Context Learning, prompt-based control (no fine-tuning) \\
\hline

~\cite{10.1145/3789514.3793205} &
Claude Sonnet 4~\cite{anthropic2025introducingclaude4}; Gemma-2-9B~\cite{google2024gemma2_9b} &
Zero-shot \ac{LLM}-based semantic negotiation using \ac{MCP}~\cite{anthropic2024mcp}(no fine-tuning) \\
\hline

~\cite{osti_10628444} &
\ac{GPT}-4~\cite{achiam2023gpt}(via the U-M \ac{GPT} service~\cite{umgpttoolkit}) &
\ac{OPRO} using structured meta-prompts (no fine-tuning) \\
\hline

~\cite{shokouhi2026agentic} &
Teacher: \ac{GPT}-5 via OpenAI \ac{API}~\cite{openai2025gpt5}; student: Qwen 2.5~\cite{hui2024qwen2} &
\ac{QLoRA}~\cite{NEURIPS2023_1feb8787} \\
\hline

~\cite{li2026multi} &
OpenAI \ac{GPT}-5~\cite{openai2025gpt5} &
Role-specific prompting with structured \ac{JSON} outputs, and \ac{RAG} over \ac{ORAN} documentation (no fine-tuning).\\
\hline

~\cite{sharma2025mobillm} &
Gemini 2.5 Flash~\cite{google_gemini25flash} &
 \ac{RAG} over MITRE FiGHT and \ac{3GPP}/\ac{ORAN} documents, with safety guardrails (no fine-tuning).\\
\hline

\cite{navidan2026toward} & 
rApp: Nvidia Nemotron~\cite{nvidia_nemotron}; 
xApp: \ac{GPT}-OSS~\cite{openai2025gptoss}; 
dApp: wireless foundation model~\cite{11264506} & 
Hierarchical control via \ac{MCP} and structured A1 policies; \acs{WPFM} fine-tuning via an \ac{MLOps}~\cite{10081336}. \\
\hline

\cite{bimo2025intent} & 
Llama-3.1~\cite{meta2024llama31_70b,touvron2023llama} & 
One-shot prompting using a five context (Instruction, Intent, Observations, Constraints, Format) and LangGraph~\cite{LangGraph} (no fine-tuning). \\ \hline

\cite{bao2025llm} &
Llama-3.1-8B~\cite{meta2024llama31_8b}&
Structured prompting for \ac{NonRT} guidance and \ac{LLM}-guided \ac{RL} training(no fine-tuning). \\
\hline

\cite{machado2026orion} & 
\ac{GPT}-5~\cite{openai2025gpt5}; Gemini 3 Pro~\cite{google2025gemini3}; Anthropic Claude Opus~\cite{anthropic2025claude45} &
\ac{MCP}-based intent translation through CAMARA (no fine-tuning) \\
\hline

\end{tabular}
\end{table*}

Li \emph{et al.}~\cite{li2026multi} present a multi-agent \ac{LLM} framework for conflict-aware rApp orchestration in \ac{ORAN}. The framework uses orchestration data, including service intents, xApp registry and profiles, active rApps/policies, retrieved \ac{ORAN} knowledge, and episodic memory of prior (intent, policy, effectiveness) tuples. It employs three \ac{LLM}-based agents: Perception, Reasoning, and Refinement, together with \ac{RAG} to transform high-level \ac{RAN} intents into deployable xApp pipelines while accounting for inter xApp and inter rApp conflicts. The paper uses \ac{RAG} to retrieve relevant \ac{ORAN} documentation and internal knowledge, which is injected into the agents' prompts to improve conflict analysis. In~\cite{10.1145/3789514.3793205}, Baena \emph{et al.} developed a semantic negotiation framework designed to solve a "who knows what" problem: applications know their own mission needs (like low battery) but not the network state, while the network knows the signal quality but not why the application needs it. The method works by using \ac{MCP} servers to expose the specific needs and tools of both the network and the application to a central \ac{LLM} planner. This planner acts like an asynchronous negotiator that queries live information, identifies a suitable compromise and creates a plan that a human operator must approve before any changes could happen. Because the system uses a query-feedback loop, \ac{LLM} can check the network again after the plan is executed to verify whether the \ac{UE}’s \ac{RSRP} improved and \ac{BLER} decreased. The used data are the following multi-domain indicators: natural-language intents, live gNodeB state (such as cell load and configuration), real-time \ac{UE} logs (including \ac{RSRP}, \ac{SINR}), and \ac{BLER}), and \ac{3GPP} technical specifications.

Machado \emph{et al.}~\cite{machado2026orion} proposes ORION, an intent-aware orchestration framework for \ac{ORAN} that integrates \ac{LLM} via \ac{MCP} to translate natural-language operator requests into enforceable network policies through a hierarchical \ac{SMO}/\ac{NonRTRIC}/\ac{NearRTRIC} pipeline. ORION combines an \ac{LLM}-assisted \ac{MCP} client/server at the \ac{SMO} for semantic translation and validation, grounded in Common \ac{API} \ac{CAMARA} \textit{Network Slice Booking} \ac{API}~\cite{CAMARA_NetworkSliceBooking, CAMARA_Project}, an rApp for \ac{SLA}-oriented policy composition, and an xApp for \ac{E2SMRC}-based enforcement at \ac{gNB}. The framework is evaluated on a dataset of 100 slice-related intents spanning \ac{eMBB}, \ac{URLLC}, and \ac{mMTC} scenarios. The \ac{LLM} in ORION follows a tool-feedback loop, using CAMARA results returned via \ac{MCP} to generate the final structured intent for rApp.

\begin{table*}[!t]
\centering
\renewcommand{\arraystretch}{1.6}
\caption{Agentic \ac{AI} Applications in \ac{ORAN}}
\label{tab:Agentic-Applicationa}

\begin{tabular}{|p{1cm}|p{2cm}|p{5.5cm}|p{1.5cm}|p{4.5cm}|}
\hline
\rowcolor{lightgray}
\multicolumn{1}{|c|}{\textbf{Ref.}} &
\multicolumn{1}{c|}{\textbf{Deployment}} &
\multicolumn{1}{c|}{\textbf{Data}} &
\multicolumn{1}{c|}{\textbf{Interface}} &
\multicolumn{1}{c|}{\textbf{Objective}} \\
\hline

~\cite{bimo2026contract} &
rApp/xApp &
Natural-language operator intents + \ac{RAN} slice telemetry (throughput, \ac{PRB}, \ac{CQI}, \acp{KPI}) &
O1 / A1 / E2 &

Safe intent-driven network slicing and \ac{SLA} assurance \\
\hline

~\cite{chatzistefanidis2025mx} &
Multi-agent \ac{LLM} &
Operator intents + O1/E2 telemetry + logs/topology/\acp{CRD} &
R1 / O1 / A1 / E2 &

Agentic observability and closed-loop \ac{RAN} reconfiguration \\
\hline

~\cite{elkael2025agentran} &
Agentic rApp/xApp/dApp hierarchy &
Natural-language operator intents + \ac{RAN} \acp{KPI}/control decisions/performance metrics + agent conversations/reasoning traces &
\ac{A2A} / \ac{MCP} / \ac{ORAN} interfaces &

Intent-driven cross-layer, cross-timescale autonomous \ac{RAN} control and self-improvement \\

\hline

~\cite{10.1145/3789514.3793205} & 
\ac{LLM} Planner + \ac{MCP} Servers & 
Natural-language intents + \ac{gNB} state + \ac{UE} logs (\ac{RSRP}/\ac{SINR})/\ac{BLER}) + application constraints + \ac{3GPP} specs & 
\ac{MCP}/\ac{A2A} &

Explainable cross-domain coordination and human-supervised proactive planning\\
\hline

~\cite{osti_10628444} &
xApp &
\ac{RAN} domain (slice-level telemetry) &
E2 &

Dynamic \ac{PRB} allocation to maximize slice utility and reliability \\
\hline

~\cite{bimo2025intent} &
\ac{SMO}/\ac{NonRT} intent-management function &
Formalized operator intents + \ac{PM} observations + configuration constraints/topology + historical strategy attempts &

O1 &

Closed-loop intent translation and energy-efficient \ac{RAN} configuration \\
\hline

~\cite{shokouhi2026agentic} &
rApp + xApps + dApp &
\ac{RAN} domain (cell-free radio state, user-rate state, \ac{ORU} state) &
A1, E2 &

Intent-driven user-weighting and \ac{ORU} activation to satisfy minimum-rate constraints and improve energy efficiency \\
\hline

~\cite{sharma2025mobillm} &
rApp (\ac{SMO}/\ac{NonRTRIC}) &
Cross-domain security data + RAN/management data &

O1, A1 &

Threat analysis, threat classification, and safe human-supervised mitigation in \ac{ORAN} \\
\hline

~\cite{navidan2026toward} & 
rApp/xApp/dApp &
Operator intents, O1/E2 telemetry, \ac{KPM} logs, Knowledge Base rules, and raw \acs{IQ}/\acs{CSI} samples & 
O1 / A1 / E2 / F1 /\ac{OFH} & 

Intent-driven slice resource control and autonomous model governance \\
\hline

~\cite{bao2025llm} &
\ac{LLM}-based rApp (\ac{NonRTRIC}) + \ac{RL}-based xApp (\ac{NearRTRIC}) &
Global multimodal O1 data (throughput trends, fault alarm logs, topology data, radio-map/digital-twin information, average channel and interference statistics) + local E2 data (channel gain, user load, expected rate state) &

O1 / A1 / E2 &

Resource management (power allocation use case) in \ac{ORAN} \\
\hline

~\cite{machado2026orion} & 
\ac{SMO} App + \ac{MCP} Client/Server + rApp + xApp &
operator intents + CAMARA slice-booking data + \ac{ORAN} policy/control context + E2 node configuration/state & 
O1 / A1 / E2 & 
Network slicing \\
\hline

\end{tabular}
\end{table*}

\subsubsection{Hierarchical Multi-Agent Agentic \ac{AI} Applications in Multiple \ac{ORAN} control Loops}
The third group represents more mature form of Agentic \ac{AI} applications in \ac{ORAN}, where agents are distributed across the \ac{NonRT}, \ac{NearRT}, and \ac{RT} layers and are assigned distinct roles, data sources, and control responsibilities.

Elkael \emph{et al.} in~\cite{elkael2025agentran} propose AgentRAN, a hierarchical Agentic \ac{AI} framework for \ac{ORAN}, in which \ac{LLM} agents translate natural-language operator intents into coordinated rApp/xApp/dApp control actions across multiple timescales, network domains, and protocol layers. Through \ac{A2A} and \ac{MCP} coordination, AgentRAN supports intent-driven control. Experiments showed that the framework can adapt resource-allocation dApp and power control dApp to changing intents without network-specific fine-tuning and can autonomously retrain an interference-prediction xApp when network conditions changed. In AgentRAN, the authors use network data (\acp{KPI}, control decisions, performance metrics, etc.) and agentic data (operator intents, inter-agent conversations, and reasoning traces). On the other hand, Shokouhi \emph{et al.}~\cite{shokouhi2026agentic} proposed a multi-agent framework for intent-driven optimization in cell-free \ac{ORAN}. A supervisor rApp translates natural-language intents into objectives, coordinating three xApps (the generated objective are sent via A1 interface): a User Weighting xApp that determines user-priority weights for the precoding dApp in the \ac{ODU}, an \ac{ORU} Management xApp for energy-saving control, and a Monitoring xApp that provides \ac{QoS} feedback and coordinates the other agents. To ensure scalability, the system uses~\cite{NEURIPS2023_1feb8787} to share a single \ac{LLM} across agents, reducing memory usage.

Navidan \emph{et al.}~\cite{navidan2026toward} presents a hierarchical agentic \ac{ORAN} architecture in which intelligence is distributed across the \ac{NonRT}, \ac{NearRT}, and \ac{RT} layers. At the strategic layer (\ac{NonRTRIC}), the \ac{LLM}-based rApp operates on high-level operator intents together with O1 measurements, \ac{KPM} logs, offline training datasets, Knowledge Base rules and operator guidelines, Model Catalog metadata, Analytics outputs, and Simulation/Digital Twin support for policy validation and model governance. At the tactical layer (\ac{NearRTRIC}), \ac{SLM} -based xApps consume the structured A1 policies from the \ac{NonRT} layer together with \ac{NearRT} E2/\ac{E2SMKPM} feedback to perform context-aware radio control and xApp orchestration. At the operational layer near the \ac{ODU}, \ac{WPFM}-based dApps operate on raw physical-layer observations, specifically \ac{IQ} samples and CSI, to provide ultra-low-latency inference for tasks such as interference detection, channel-related decisions, and other \ac{PHY}-level actions.

Overall, these works showed that Agentic \ac{AI} in \ac{ORAN} is evolving toward a hierarchical architecture in which different agents consume different forms of data and coordinate different control loops in a whole \textit{autonomous control} of \ac{ORAN}.

\section{Challenge view of \ac{ORAN} Data}\label{sec:Challenges}

\subsection{Data Security}
 \ac{ORAN} data security is challenging as the same architectural properties that enable openness, disaggregation, virtualization, and intelligent control also enlarge the attack surface across open interfaces, multi-vendor components, cloud platforms, and third party xApps/rApps/dApps~\cite{10467183,liang2026security, KARBALAEEMOTALLEB2025, toward-ojcoms}. As a result, \ac{ORAN} telemetry, control, policy, and management data are exposed to threats against their confidentiality, integrity, and availability. The problem is especially severe for \ac{AIML} \ac{ORAN} applications, because malicious manipulation of training or inference data can mislead the \ac{NearRTRIC} or \ac{NonRTRIC} into issuing harmful control decisions~\cite{luo2025attack}. Security threats span both conventional interface level attacks, such as spoofing, \ac{MITM} attacks, malicious xApps, or unauthorized configuration, and \ac{AI} level attacks, such as data poisoning, evasion, backdoor insertion, or policy infiltration~\cite{10467183,11028721,11106439,luo2025attack,LACAVA2025111727}. We next summarize examples of security threats on \ac{ORAN} data:

\begin{itemize}
 \item \textit{\ac{KPM} poisoning:} False \acp{KPM} can be injected into \ac{NearRT} control loops, through spoofing on E2 interface or compromised E2 nodes, corrupting the data consumed by xApps. A possible mitigation is detection of anomalous \acp{KPM} values before xApps use them~\cite{mehrban2025integrating}. As~\cite{11028721} proposes an \ac{LSTM} detector for poisoned \acp{KPM}.

 \item \textit{\ac{UE} metrics poisoning:} Attackers may manipulate reported \ac{UE} metrics, for example through botnets of compromised \acp{UE}, to poison the data used by \ac{ML} models in \ac{NearRT} and \ac{NonRTRIC}~\cite{11106439}. A mitigation example is the security pipeline proposed in~\cite{11106439}, which combines: UE risk analysis xApp, misbehavior detection xApp, and resilience management rApp.

 \item \textit{Training-data poisoning and inference-time evasion:} Poisoning alters training samples, whereas evasion manipulates runtime or evaluation inputs after training; both can degrade the predictions used for \ac{ORAN} optimization. Moreover, Luo-Chen \emph{et al.}~\cite{luo2025attack} show that poisoned data can affect not only prediction accuracy, but also \ac{ML} pipeline steps such as feature selection and hyperparameter optimization.

 \item \textit{Backdoor attacks against xApps:} Poisoned training datasets can implant hidden triggers into \ac{MLDL} xApps, so that xApps behave normally under ordinary conditions but produce harmful control actions when a trigger appears~\cite{LACAVA2025111727}. This highlights the need to secure the full xApps training pipeline, not only runtime inference.

 \item \textit{Adversarial manipulation of \ac{AI} inputs:} Beyond explicit poisoning of training data, attackers may manipulate the online inputs used by \ac{AI} control models, causing incorrect scheduling, allocation, or policy decisions. This reflects the vulnerability of \ac{AI} \ac{ORAN} control loops to adversarial or corrupted inputs values~\cite{luo2025attack,benzaid2025ai}.

 \item \textit{Malicious applications and misconfiguration:} \ac{ORAN} data can also be compromised through malicious xApps/rApps, and unsafe or inconsistent configuration changes across disaggregated components. Such problems may alter the collection, forwarding, or interpretation of data and can propagate incorrect control decisions~\cite{10467183,liang2026security,KARBALAEEMOTALLEB2025, toward-ojcoms}. Runtime attestation has therefore been proposed to verify the integrity of deployed xApps/rApps against trusted reference measurements and detect unauthorized modifications during operation~\cite{attestation}.

 \item \textit{Unauthorized access and \ac{MITM} attacks:} Since \ac{ORAN} relies on open interfaces and distributed components, attackers may target interfaces such as E2~\cite{toward-ojcoms} or \ac{OFH} to tamper or inject data if authentication, integrity protection, and auditing are weak~\cite{10467183}. This motivates strong interface protection~\cite{liang2026security,KARBALAEEMOTALLEB2025}.

\end{itemize}

\subsection{Data Privacy}

Data privacy in \ac{ORAN} is challenged by \ac{ORAN} architecture itself: openness and multi-vendor. Unlike traditional \ac{RAN} systems where operational data remain within a single vendor domain, \ac{ORAN} enables third-party xApps, rApps, and dApps to access \ac{ORAN} data. We next summarize examples of privacy threats on \ac{ORAN} data: 

\begin{itemize}

\item \textit{Malicious or misconfigured \acp{RIC} components:} 
In multi-stakeholder environments, malicious or misconfigured applications~\cite{yungaicela2024misconfiguration} may infer sensitive information from shared \ac{RIC} databases (The \ac{NearRTRIC} database stores \ac{RAN} and \ac{UE} telemetry consumed by xApps, including traffic patterns, service usage, and mobility characteristics). Moreover, the risks associated with storing spectrogram data in shared \ac{ORAN} database are highlighted in~\cite{10632857}, which shows that \ac{MLDL} xApps operating on shared databases can expose sensitive wireless data without adequate protection. Furthermore, internal \ac{RIC} components may attempt unauthorized access to databases. The Zero-Trust \ac{RIC} framework in~\cite{10976081} mitigates this risk by encrypting data prior to storage and enabling privacy-preserving inference over protected data, thereby limiting both external interception and insider misuse.

\item \textit{Open Interfaces:} 
Open interfaces further increase data exposure in \ac{ORAN}. Interfaces such as E2, O1, A1, and O2 can be exploited through several attacks, including \ac{MITM}, data tampering, and unauthorized access~\cite{bib2,298957}. Such attacks may lead to leakage of \ac{UE} data, or inference of mobility or traffic patterns. Moreover, the disaggregated of \ac{RIC} components across cloud and edges increases data exposure risks, especially when telemetry traverses heterogeneous administrative boundaries~\cite{10329927}. Therefore, securing open interfaces requires end-to-end encryption, strong mutual authentication, integrity protection, and fine-grained authorization policies.

\item \textit{Virtualization:} 
The transition toward virtualization and cloud-native architectures in \ac{ORAN} introduces additional trust boundaries across \ac{OCU}, \ac{ODU}, \ac{ORU}, and \ac{RIC} components~\cite{10872859}. Virtualization and containerization enlarge the privacy attack surface, necessitating system-level safeguards such as encryption, or access control.

\end{itemize}

Privacy risks are amplified in \ac{AI} based \ac{ORAN} applications. \ac{AI} models may leak sensitive data through: 
(i) \textit{membership inference}, where an adversary determines whether a specific sample was part of the training dataset~\cite{9706362}; 
(ii) \textit{model inversion}, where input features are reconstructed from model outputs~\cite{10.1145/3789456.3789473}; 
(iii) \textit{gradient leakage}, where shared gradients reveal properties of local training data; and 
(iv) \textit{data reconstruction attacks}, where original training data are recovered from model parameters or intermediate representations~\cite{10.1145/3789456.3789473}.

Moreover, \ac{FL} introduces new privacy threats such as malicious model updates. The integrated Zero-Trust and privacy-preserving \ac{FL} framework in~\cite{fi17060233} demonstrates that differential privacy, secure aggregation, and continuous trust enforcement can be used to protect \ac{FL} in \ac{ORAN}.

\subsection{Data Scarcity}
\label{sec:data_scarcity}
The high performance of \ac{ORAN} management applications, especially \ac{AIML} based ones, depends on the availability of large-scale and diverse training data. However, data scarcity remains a fundamental challenge, and arises from:
\begin{itemize}
 \item \textit{Lack of datasets}: although \ac{ORAN} promotes openness and interoperability, management and telemetry data collected via O1/O2 interfaces and \acp{RIC} remain restricted to the operator use. However, no large, scale public \ac{ORAN} datasets currently exist~\cite{10604823}.
 
 \item \textit{Restricted exploration}: operators typically explore only a restricted set of configuration parameters in live networks to avoid service degradation, resulting in incomplete representation of cases~\cite{10113782}. This leads to both insufficient samples and insufficient diversity.
 
 \item \textit{Evolving traffic}: Models trained on historical network traffic traces rapidly degrade under distribution shifts caused by new services, encryption methods or adversarial behavior. As shown in recent \ac{ORAN} digital twin study~\cite{11224872}, rare events such as sudden latency spikes, or abnormal \acp{KPM} are underrepresented in real traffic. 
 
 \item \textit{Under-representation}: data for rare and anomalies scenarios are inherently scarce~\cite{11434085}. This further restricts learning approaches in \ac{RAN} management.
 
 \item \textit{Privacy \textbf{constrains}}: \ac{ORAN} data may expose sensitive user information and operator configurations~\cite{10976081}. Regulatory requirements, such as \ac{GDPR}~\cite{GDPR} and commercial confidentiality limit data sharing. Consequently, datasets are often anonymized or aggregated, reducing granularity and constraining the effectiveness of \ac{AIML} model training.
 
\end{itemize}

Addressing the scarcity challenge requires hybrid strategies, including synthetic data generation~\cite{11152840, amachaghi2025improving}, digital twins~\cite{10643670}, transfer learning~\cite{9388790}, few-shot learning~\cite{10.1145/3386252}, simulator/testbed-driven data collection~\cite{10.1109/DySPAN53946.2021.9677430}, and privacy-preserving data sharing mechanisms~\cite{10113782,11224872}. Without mitigation of data scarcity, \ac{AI}/agentic \ac{ORAN} automation risks unreliable decision-making.

\subsection{Data Heterogeneity}
\ac{ORAN} data are inherently heterogeneous because it is generated by multiple disaggregated component (\ac{ORU}, \ac{ODU}, \ac{OCU}, \acp{RIC}, \ac{SMO}, O-Cloud, and external sources). These components expose diverse data types, such as packet traces, radio measurements, \acp{KPM}, configuration parameters, logs, control messages, mobility events, and user/operator intent. Consequently, \ac{ORAN} data differs in format, semantics, granularity, sampling rate, and abstraction level~\cite{10024837,10329915,10439167}. 
Moreover, this challenge is amplified in multi-vendor deployments, where similar measurements may be exposed through different data models, naming conventions, reporting periods, and implementation interpretations. Therefore, combining data across \ac{ORAN} domains requires semantic alignment, synchronization, and normalization. For \ac{AIML}- applications, such heterogeneity can lead to \ac{NONIID} training data, and poor model transferability, and generalizability~\cite{feng2026ai,11062617, bermudez2025graph}.

\subsection{Data Quality}
 The effectiveness of \ac{ORAN} applications is strongly influenced by the quality of the collected data. Quality challenges are related to noise, timeliness, and consistency.

\subsubsection{Noisy Data}
\ac{ORAN} data are subject to various sources of noise. For example, radio measurements and \acp{KPM} may exhibit fluctuations caused by interference, mobility, and measurement inaccuracy. Such noise can distort \ac{AI} models input and degrade their performance~\cite{11224872}.

\subsubsection{Data Timeliness}
\ac{ORAN} operates across multiple control loops with different latency requirements. Delays in data collection can lead to old data being used for decision making. Yuan \emph{et al.}~\cite{11103489} highlight that effective data lifecycle management is required to ensure data timeliness.

\subsubsection{Data Consistency} is challenged by the multi-vendor natural of \ac{ORAN}. Differences in data formats, \ac{KPM} definitions, and reporting intervals may lead to conflicting data. As discussed in~\cite{10969847}, the lack of unified data models and interoperability across \ac{ORAN} components introduces inconsistencies that complicate data fusion and reduce the reliability of \ac{AI} applications.

\subsection{Data Preprocessing} 
Data Preprocessing is a necessary step for enabling reliable \ac{ORAN} applications, especially \ac{AI} ones. Raw \ac{ORAN} data are collected from different components, which makes it heterogeneous, noisy, and imbalanced. Thus, it cannot be directly used for analysis or model training. Preprocessing focuses on improving data quality before model development~\cite{10.1145/3711118}. Typical preprocessing steps include: (i) \textit{data cleaning}, by removing noise and missing values; (ii) \textit{data transformation}, by converting heterogeneous data into a consistent format, including normalization and encoding; (iii) \textit{feature extraction}, by deriving useful features from raw measurements; (iv) \textit{temporal alignment}, by synchronizing data collected at different timescales to ensure correct correlation across components; (v) \textit{data aggregation}, by reducing data volume through summarizing information over time windows or flows; and (vi) \textit{data segmentation}, by dividing time-series data into windows.

In \ac{ORAN}, preprocessing is integrated with data collection and analysis pipelines. For example, testbed-based \ac{ORAN} implementations apply filtering, segmentation, and feature extraction to telemetry and log data before feeding it into xApps for decision-making~\cite{kumar2026intelligent,10.1145/3737900.3770164, ETSI-TR104037}. Despite the importance of data preprocessing in \ac{ORAN}, it remains challenging due to the lack of standardized data formats, the diversity of vendor implementations, and the need for low-latency processing~\cite{nGRG_RR_2024_10_dApps}. These limitations motivate the need for automated and standardized preprocessing pipelines for future \ac{ORAN} deployments.

\section{Conclusions} \label{sec:conclusion}
This paper presented a comprehensive data-centric overview of \ac{ORAN} applications. We first introduced the main \ac{ORAN} architectural components. Then, we proposed a data-centric framework that organizes \ac{ORAN} data according to five views: source, accessibility, abstraction level, application usage, and challenges. Based on this framework, we reviewed the main \ac{ORAN} data types, including radio-level data, \ac{RAN} telemetry, \acp{KPM}, management and historical data, control-plane signaling, cross-domain data, intelligence-derived data, and user intent data. We then discussed how these data are exposed through \ac{ORAN}, \ac{3GPP} and Agentic \ac{AI} interfaces, and how they are transformed from raw measurements into aggregated, derived, enriched, and intelligence-derived information. The paper also mapped these data types to statistical, \ac{AIML}-based, and Agentic \ac{AI} applications deployed as rApps, xApps, dApps, and hierarchical control frameworks.

Finally, we highlighted key challenges, including data scarcity, heterogeneity, quality, timeliness, privacy, security, and trustworthiness. We believe that the proposed data-centric framework can support future research toward more reliable, scalable, and autonomous \ac{ORAN} intelligence. Future work should focus on standardized data exposure, reusable preprocessing pipelines, trustworthy and privacy-preserving data sharing, and safe Agentic \ac{AI} control loops.

Several future research directions emerge from challenges identified in this survey. First, standardized and semantically aligned data models are needed to support consistent data exposure across \ac{ORAN}, \ac{3GPP}, and emerging Agentic \ac{AI} interfaces. Such models should define common data formats and semantics and include metadata related to data provenance and timestamps. Second, representative \ac{ORAN} datasets are needed to combine various types of data: radio measurements, \ac{RAN} telemetry, management data, logs, and operator intents. Synthetic data, digital twins, testbeds, and data-sharing mechanisms can help address data scarcity. Third, automated and low latency preprocessing and data quality assessment pipelines should support data cleaning, normalization, temporal alignment, and the detection of noisy, missing, delayed, or inconsistent data across different control timescales. Finally, future intelligent control loops should be resilient to poisoned or manipulated data and should validate the quality, timeliness, and confidence of their inputs. Human supervision and post action monitoring should also be considered for high-impact control operations. These directions should be evaluated in real testbeds to determine whether intelligent \ac{ORAN} applications remain reliable, and secure under different network conditions.

\section*{Acknowledgment} 
This research was partially supported by the German Federal Ministry of Education and Research (BMBF) through the Open6GHub+ project (Grant No. $\mathrm{16KIS2402K}$) and the KIOps6G project (Grant No. $\mathrm{16KIS2398}$), partially funded by the European Commission through the Agentic-6G project (Grant No. $\mathrm{101290342}$), and partially supported by UKRI (UK ID: 640240), CDTI (ES), ANI (PT ID: COMPETE2030-FEDER-01602800), and TÜBİTAK (TR ID: 9240013) within the EUREKA CELTIC-NEXT 6G-SMART project (C2023/2-20).

\bibliographystyle{IEEEtran}
\bibliography{refereces} 

\end{document}